\documentclass[sigplan,copyrightmode=0]{acmart}

\def\BibTeX{{\rm B\kern-.05em{\sc i\kern-.025em b}\kern-.08emT\kern-.1667em\lower.7ex\hbox{E}\kern-.125emX}}

\usepackage{wrapfig}
\usepackage{multirow}
\usepackage{balance}

\usepackage{pifont}

\newcommand{\smallcapital}{\fontsize{9pt}{10pt}\selectfont}

\newcommand{\skiourakisize}{\fontsize{8pt}{10pt}\selectfont}

\usepackage{verbatim}   % allows multi-line comments

\usepackage{rotating}

\usepackage{arydshln}

\usepackage{placeins}

\begin{document}

%
% The "title" command has an optional parameter, allowing the author to define a "short title" to be used in page headers.
\title{Leveraging Deep Learning to Improve the Performance Predictability of Cloud Microservices}

%
% The "author" command and its associated commands are used to define the authors and their affiliations.
% Of note is the shared affiliation of the first two authors, and the "authornote" and "authornotemark" commands
% used to denote shared contribution to the research.
\author{Yu Gan}
\affiliation{%
  \institution{Cornell University}
  }
\email{yg397@cornell.edu}

\author{Yanqi Zhang}
\affiliation{%
  \institution{Cornell University}
}
\email{yz2297@cornell.edu}

\author{Kelvin Hu}
\affiliation{%
  \institution{Cornell University}
  }
\email{sh2442@cornell.edu}

\author{Dailun Cheng}
\affiliation{%
  \institution{Cornell University}
  }
\email{dc924@cornell.edu}

\author{Yuan He}
\affiliation{%
  \institution{Cornell University}
  }
\email{yh772@cornell.edu}

\author{Meghna Pancholi}
\affiliation{%
  \institution{Cornell University}
  }
\email{mp832@cornell.edu}

\author{Christina Delimitrou}
\affiliation{%
  \institution{Cornell University}
  }
\email{delimitrou@cornell.edu}

\begin{abstract}
	Performance unpredictability is a major roadblock towards cloud adoption, and has performance, cost, and revenue ramifications. 
Predictable performance is even more critical as cloud services transition from monolithic designs to microservices. 
Detecting QoS violations after they occur in systems with microservices results in long recovery times, as hotspots 
propagate and amplify across dependent services. % causing cascading quality-of-service violations. 
%does not avoid the degraded 
%performance to begin with. This is even more detrimental as cloud applications increasingly move away from traditional monolithic services, 
%and instead consist of hundreds of loosely-coupled microservices. Dependencies between microservices can result in cascading QoS violations, 
%amplifying unpredictable performance through the system. 

We present Seer, an online cloud performance debugging system that leverages 
deep learning and the massive amount of tracing data cloud systems collect to learn spatial and temporal %usage 
patterns that translate to QoS violations. Seer combines lightweight 
distributed RPC-level tracing, with detailed low-level hardware monitoring 
to signal an upcoming QoS violation, and diagnose the source of unpredictable 
performance. Once an imminent QoS violation is detected, Seer notifies 
the cluster manager to take action to avoid performance degradation 
altogether. We evaluate Seer both in local clusters, and in large-scale 
deployments of end-to-end applications built with microservices with hundreds 
of users. We show that Seer correctly anticipates QoS violations 91\% of the time, 
and avoids the QoS violation to begin with in 84\% of cases. Finally, 
we show that Seer can identify application-level design bugs, 
and provide insights on how to better architect microservices to achieve predictable performance. 
%when violations are detected early, corrective actions are almost always successful in avoiding them. 
%the scheduler is successful in avoiding them altogether. 
%corrective action to be applied successfully. 

\vspace{3in}

\end{abstract}

\maketitle

\section{Introduction}

Cloud computing services are governed by strict quality of service (QoS) constraints 
in terms of throughput, and more critically tail latency~\cite{BarrosoBook,tailatscale,Delimitrou13,Delimitrou14}. Violating 
these requirements worsens the end user experience, leads to loss of availability and 
reliability, and has severe revenue implications~\cite{BarrosoBook, Barroso11, tailatscale,Delimitrou16,Delimitrou17,Delimitrou13d}. 
In an effort to meet these performance constraints and facilitate frequent application updates, cloud services have recently undergone 
a major shift from complex monolithic designs, which encompass the entire functionality in a single binary, 
to graphs of hundreds of loosely-coupled, single-concerned microservices~\cite{Cockroft16,Gan19}. 
Microservices are appealing for several reasons, including accelerating development and deployment, simplifying correctness 
debugging, as errors can be isolated in specific tiers, and enabling a rich software ecosystem, as each microservice is written 
in the language or programming framework that best suits its needs. %heterogeneity %requirements

At the same time microservices signal a fundamental departure from the way traditional cloud applications were designed, and bring with them 
several system challenges. Specifically, even though the quality-of-service (QoS) requirements of the end-to-end application are similar for microservices 
and monoliths, the tail latency required for each individual microservice is much stricter than for traditional cloud applications~\cite{Meisner11,Lo14,Lo15,Delimitrou14,GoogleTrace,agile,Mars13b,Gan18,Gan18b,Gan19}. 
This puts increased pressure on delivering predictable performance, as dependencies between microservices mean that a single misbehaving microservice 
can cause cascading QoS violations across the system. %, as a single mismanaged microservice backpressures its neighboring services. 

\begin{wrapfigure}[17]{r}{0.245\textwidth}
	\vspace{-0.2in}
	\centering
	\includegraphics[scale=0.344, trim=1.6cm 1.8cm 5.6cm 5.8cm, clip=true]{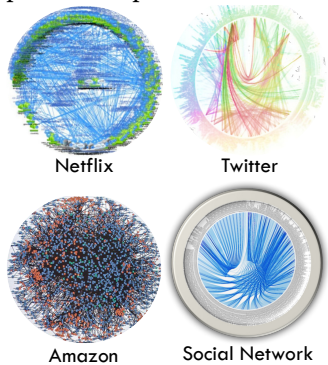}
	\caption{{\bf{\label{fig:deathstars}}} {Microservices graphs in three large 
	cloud providers~\cite{Cockroft15,Cockroft16,twitter_decomposing}, and our \texttt{Social Network} service. }} 
\end{wrapfigure}

Fig.~\ref{fig:deathstars} shows three instances of real large-scale production deployments of microservices~\cite{Cockroft15,Cockroft16,twitter_decomposing}. 
The perimeter of the circle (or sphere surface) shows
the different microservices, and edges show dependencies between them. We also show these dependencies for \textit{Social Network}, one of the large-scale services 
used in the evaluation of this work (see Sec.~\ref{sec:methodology}). %\note{missing something}
Unfortunately the complexity of modern cloud services means that manually determining the impact of each pair-wide dependency on end-to-end QoS, or relying on the user to provide this information is impractical. %prohibitively expensive. 

Apart from software heterogeneity, datacenter hardware is also becoming increasingly heterogeneous as 
special-purpose architectures~\cite{tpu,diannao,shidiannao,diannao_family,dadiannao} 
and FPGAs are used to accelerate critical operations~\cite{catapult,catapult2,brainwave,firestone18}. This adds to the existing server heterogeneity in the cloud 
where servers are progressively replaced and upgraded over the datacenter's provisioned 
lifetime~\cite{Mars13a,mars11b,Delimitrou13,Delimitrou14b,Nathuji07}, and further complicates the effort to guarantee
predictable performance. 

The need for performance predictability has prompted a long line of work on performance tracing, monitoring, and debugging systems~\cite{dapper,gwp,MysteryMachine,xtrace,cloudseer,Xu16,nsdi18}. 
Systems like Dapper and {\smallcapital GWP}, for example, rely on distributed tracing (often at {\smallcapital RPC} level) and low-level hardware event monitoring respectively to detect performance abnormalities, 
while the Mystery Machine~\cite{MysteryMachine} leverages the large amount of logged data 
to extract the causal relationships between requests. %, and {\smallcapital HUYGENS} uses tracing to synchronize the clocks of a large cluster~\cite{nsdi18}. 

\begin{figure}
\centering
\begin{tabular}{ccc}
	\includegraphics[scale=0.20, viewport=140 10 300 65]{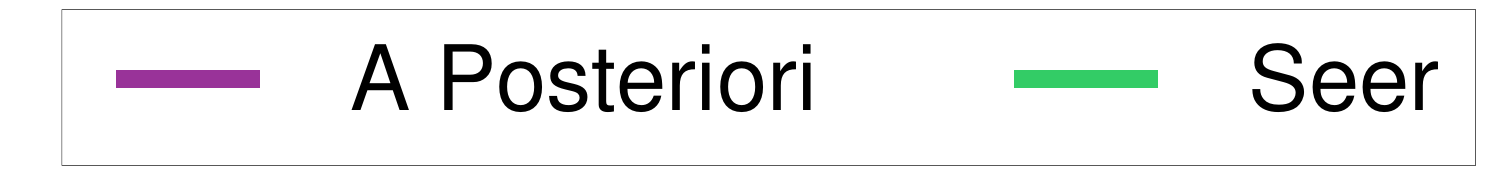} &  & \\
\includegraphics[scale=0.165, viewport=80 0 400 440]{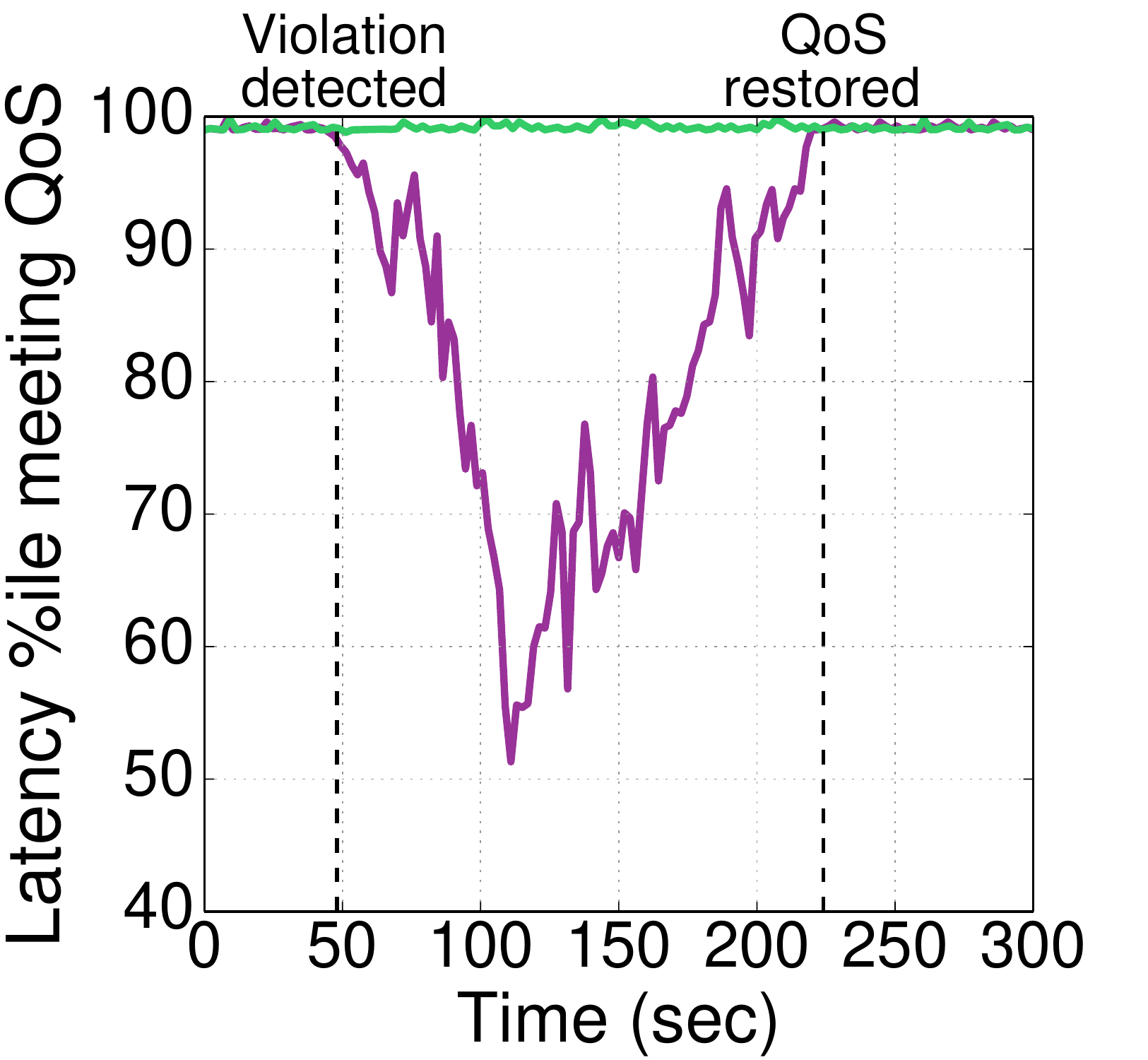} & 
\includegraphics[scale=0.165, viewport=20 0 440 440]{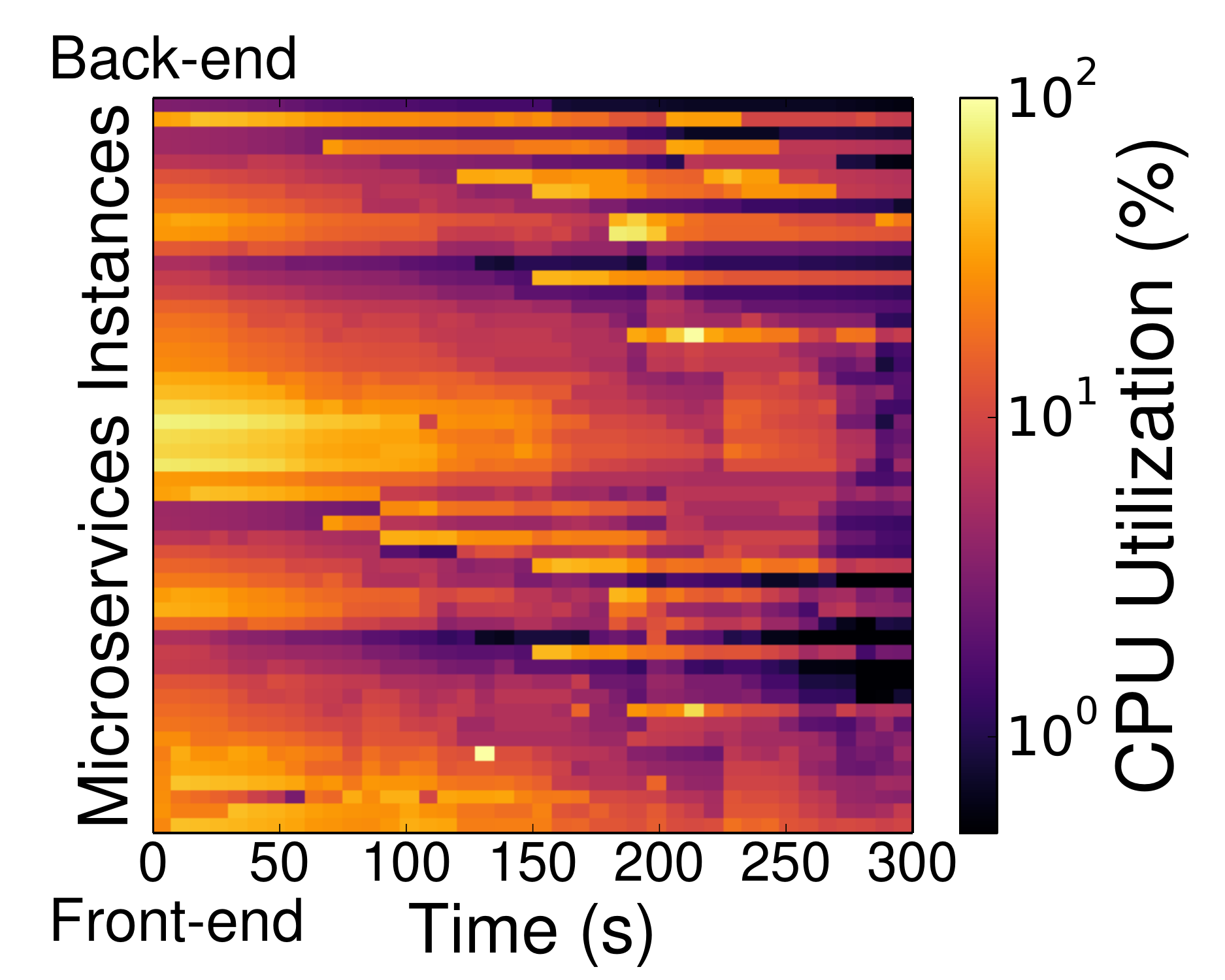} & 
\includegraphics[scale=0.165, viewport=-20 0 400 440]{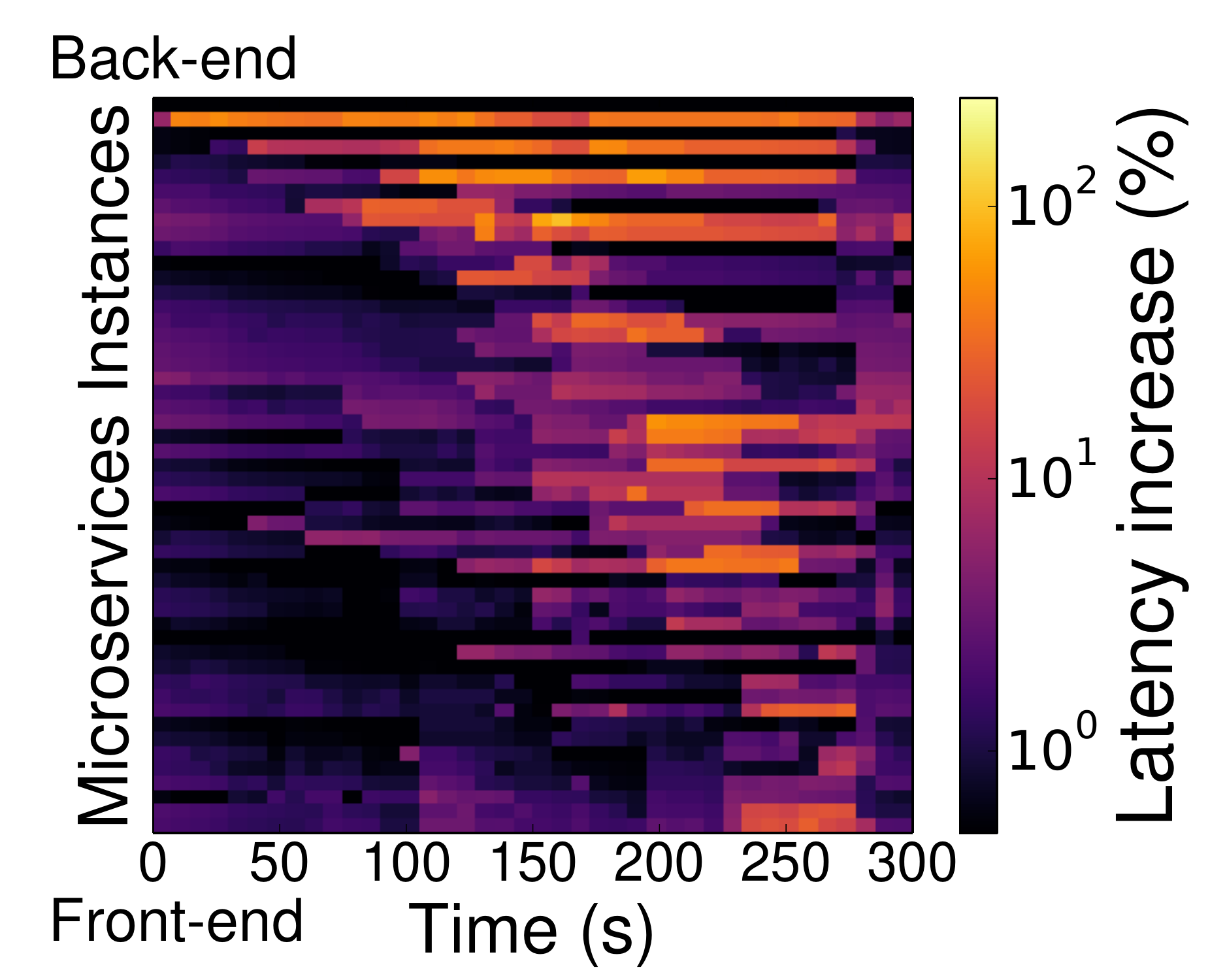}
\end{tabular}
\caption{\label{fig:motivation} {The performance impact of a posteriori performance diagnostics for a monolith and for microservices. }}
\end{figure}

Even though such tracing systems help cloud providers detect QoS violations and 
apply corrective actions to restore performance, until those actions take effect, performance suffers. 
For monolithic services this primarily affects the service experiencing the QoS violation itself, 
and potentially services it is sharing physical resources with. 
With microservices, however, a posteriori QoS violation detection is more impactful, 
as hotspots propagate and amplify across dependent services, 
forcing the system to operate in a degraded state for longer, until all oversubscribed tiers 
have been relieved, and all accumulated queues have drained. %The longer the system stays in this oversubscribed 
Fig.~\ref{fig:motivation}a shows the impact of reacting to a QoS violation after it occurs 
for the \textit{Social Network} application with several hundred users running on 20 two-socket, high-end servers. 
Even though the scheduler scales out all oversubscribed tiers once the violation occurs, 
it takes several seconds for the service to return to nominal operation. There are two reasons for this; 
first, by the time one tier has been upsized, its neighboring tiers have built up request backlogs, which cause them to saturate in turn. 
Second, utilization is not always a good proxy for tail latency and/or 
QoS violations~\cite{Lo14,Lo15,BarrosoBook,tailatscale,Ousterhout13}. Fig.~\ref{fig:motivation}b shows the utilization 
of all microservices ordered from the back-end to the front-end over time, 
and Fig.~\ref{fig:motivation}c shows their corresponding 99$^{th}$ percentile 
latencies normalized to nominal operation. 
Although there are cases where high utilization and high latency match, 
the effect of hotspots propagating through the service is much more pronounced when looking at latencies, 
with the back-end tiers progressively saturating the service's logic and front-end microservices. 
In contrast, there are highly-utilized microservices that do not experience increases in their tail latency. 
A common way to address such QoS violations is rate limiting~\cite{Suresh17}, 
which constrains the incoming load, until hotspots dissipate. This restores performance, but degrades the end user's experience, as a fraction of input requests is dropped. 

We present \textit{Seer}, a proactive cloud performance debugging system that leverages practical deep learning techniques 
to diagnose upcoming QoS violations in a scalable and online manner. First, Seer is proactive to avoid the long recovery periods of 
a posteriori QoS violation detection. Second, it uses the massive amount of tracing data cloud systems collect over time to learn 
spatial and temporal patterns that lead to QoS violations early enough to avoid them altogether. Seer includes a lightweight, 
distributed {\smallcapital RPC}-level tracing system, based on Apache Thrift's timing interface~\cite{thrift}, to collect end-to-end traces of request execution, 
and track per-microservice outstanding requests. 
Seer uses these traces to train a deep neural network to recognize imminent QoS violations, 
and identify the microservice(s) that initiated the performance degradation. 
Once Seer identifies the culprit of a QoS violation that will occur over the next few 100s of milliseconds, it uses detailed per-node hardware monitoring 
to determine the reason behind the degraded performance, and provide the cluster scheduler 
with recommendations on actions required to avoid it. 

We evaluate Seer both in our local cluster of 20 two-socket servers, and on large-scale clusters on Google Compute Engine (GCE) with a set 
of end-to-end interactive applications built with microservices, including the \textit{Social Network} above. 
In our local cluster, Seer correctly identifies upcoming QoS violations in 93\% of cases, 
and correctly pinpoints the microservice initiating the violation 89\% of the time. To combat long inference times as clusters scale, 
we offload the {\smallcapital DNN} training and inference to Google's Tensor Processing Units ({\smallcapital TPU}s) when running on {\smallcapital GCE}~\cite{tpu}. We additionally experiment with 
using {\smallcapital FPGA}s in Seer via Project Brainwave~\cite{brainwave} when running on Windows Azure, and show that both types of acceleration speed up Seer by 200-235x, 
with the {\smallcapital TPU} helping the most during training, and vice versa for inference. Accuracy is consistent with the small cluster results. 

Finally, we deploy Seer in a large-scale installation of the \textit{Social Network} service with several hundred users, and show that it not only 
correctly identifies 90.6\% of upcoming QoS violations and avoids 84\% of them, but that detecting patterns that create hotspots helps the application's 
developers improve the service design, resulting in a decreasing number of QoS violations over time. As cloud application and hardware complexity continues 
to grow, data-driven systems like Seer can offer practical solutions for systems whose scale make empirical approaches intractable. 

%~\cite{BarrosoBook}

\section{Related Work}
\label{sec:RelatedWork}

\begin{table*}
\skiourakisize
\centering
\begin{tabular}{ccccc}
\hline
\multirow{2}{*}{\bf Service} & {\bf{Communication}} & {\bf{Unique}} & {\bf{Per-language LoC breakdown}} \\
			     & {\bf{Protocol}} & {\bf{Microservices}} & {\bf{(end-to-end service)}} \\
\hline
{\bf{Social Network}} & {\bf RPC} & {\texttt{36}} & {34\% C, 23\% C++, 18\% Java, 7\% node, 6\% Python, 5\% Scala, 3\% PHP, 2\% JS, 2\% Go} \\
%{\bf{Network}} & & & {6\% Python, 5\% Scala, 3\% PHP, 2\% Javascript, 2\% Go} \\
\hdashline[0.5pt/2.5pt]
{\bf{Media Service}} & {\bf RPC} & {\texttt{38}} & {30\% C, 21\% C++, 20\% Java, 10\% PHP, 8\% Scala, 5\% node, 3\% Python, 3\% JS} \\
%{\bf{Reviewing}} & & & {8\% Scala, 5\% node.js, 3\% Python, 3\% Javascript} \\
\hdashline[0.5pt/2.5pt]
{\bf{E-commerce Site}} & {\bf REST} & {\texttt{41}} & {21\% Java, 16\% C++, 15\% C, 14\% Go, 10\% JS, 7\% node, 5\% Scala, 4\% HTML, 3\% Ruby} \\
%{\bf{Website}} & {\bf{RPC}} & & {7\% node.js, 5\% Scala, 4\% HTML, 3\% Ruby} \\
\hdashline[0.5pt/2.5pt]
{\bf{Banking System}} & {\bf{RPC}} & {\texttt{28}} & {29\% C, 25\% Javascript, 16\% Java, 16\% node.js, 11\% C++, 3\% Python} \\
%{\bf{System}} & & & {16\% node.js, 11\% C++, 3\% Python} \\
\hdashline[0.5pt/2.5pt]
{\bf{Hotel Reservations~\cite{gomicroservices}}} & {\bf{RPC}} & {\texttt{15}} & {89\% Go, 7\% HTML, 4\% Python} \\
%\hdashline[0.5pt/2.5pt]
%{\bf{}} & \multirow{2}{*}{\bf{RPC}} & \multirow{2}{*}{\texttt{7}} & {89\% Go, 7\% HTML, 4\% Python} \\
\hline
\end{tabular}
\caption{\label{loc_stats} {Characteristics and code composition of each end-to-end microservices-based application. } } % (at time of paper submission). }
\end{table*}

\begin{figure*}%[18]{l}[\dimexpr\columnwidth+\columnsep\relax]{11cm}
\centering
\begin{minipage}{0.58\textwidth}
\centering
\includegraphics[scale=0.39, trim=0 0cm 0 4cm, clip=true]{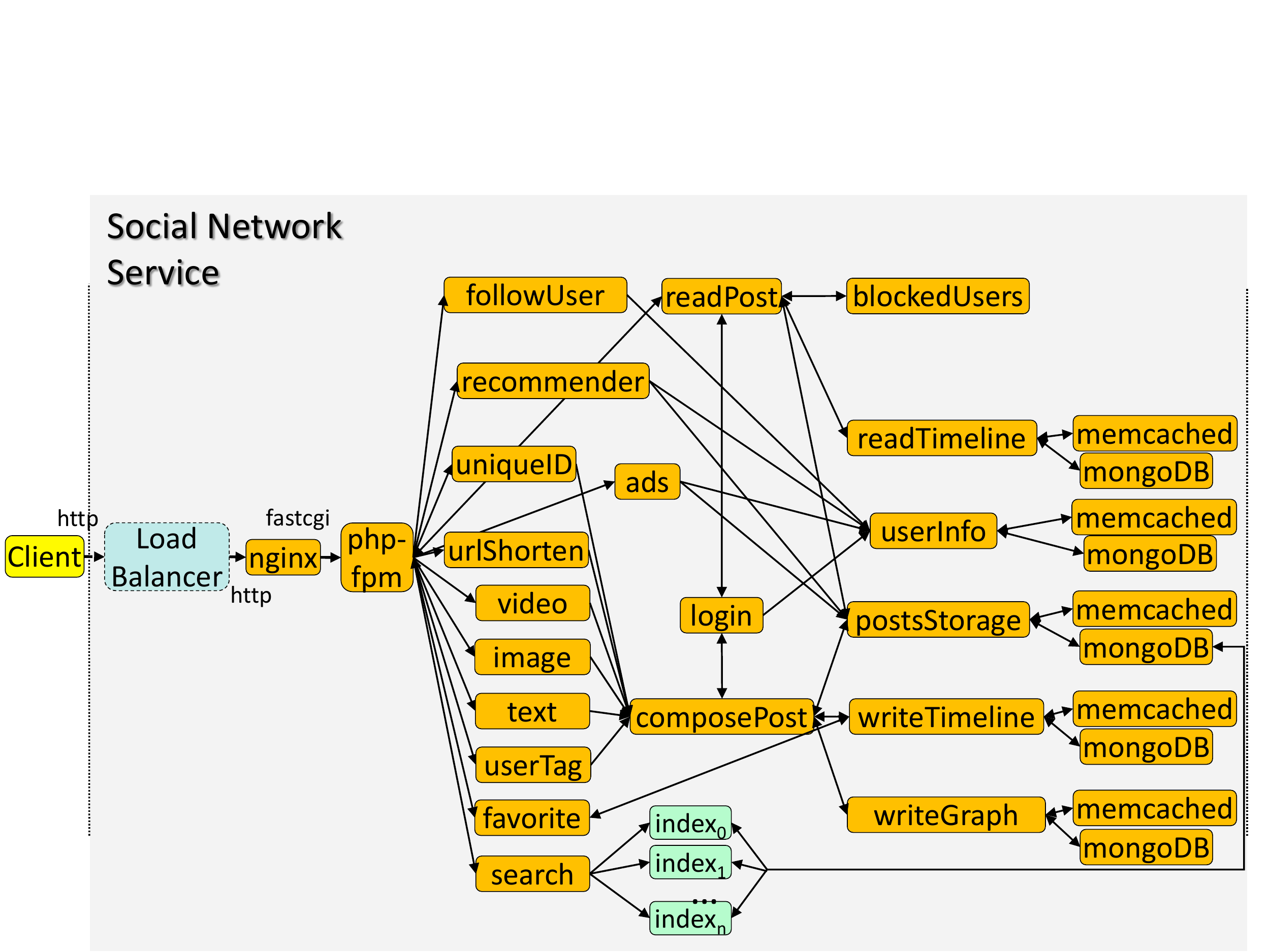}
\caption{\label{fig:social} {Dependency graph between the microservices of the end-to-end \textit{Social Network} application. }}
\end{minipage}
\hspace{0.6cm}
\begin{minipage}{0.36\textwidth}
\includegraphics[scale=0.39, trim=0cm 0cm 0cm 4cm, clip=true]{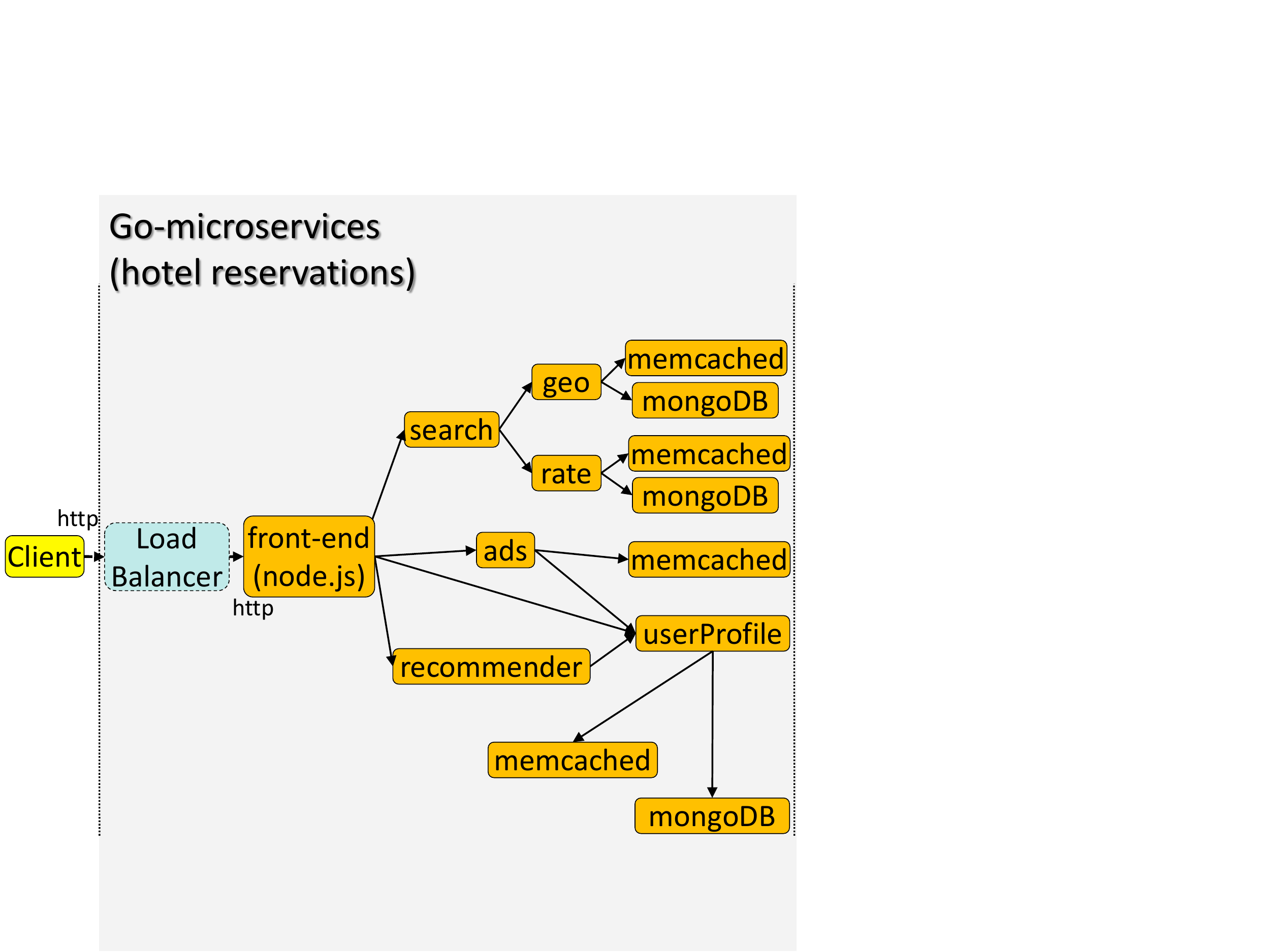}
\caption{\label{fig:hotel} Architecture of the hotel reservation site using Go-microservices~\cite{gomicroservices}. }
\end{minipage}
\end{figure*}

Performance unpredictability is a well-studied problem in public clouds that stems from 
platform heterogeneity, resource interference, software bugs and 
load variation~\cite{Cherkasova07,Mangot09,Schad10,Delimitrou14, Delimitrou13d, Delimitrou16, Iosup10, Iosup11, Delimitrou15, Schad10, Rehman10, Khamra10,Lo14,Lo15}. 
We now review related work on reducing performance unpredictability in cloud systems, including through scheduling and cluster management, or through online tracing systems. 

\vspace{0.05in}
\noindent{\bf{Cloud management: }}The prevalence of cloud computing has motivated several cluster management designs. Systems like Mesos~\cite{Mesos11},
Torque~\cite{torque}, Tarcil~\cite{Delimitrou15}, and Omega~\cite{omega13} all target the problem of resource allocation
in large, multi-tenant clusters. 
Mesos is a two-level scheduler. It has a central coordinator that makes resource
offers to application frameworks, and each framework has an individual scheduler
that handles its assigned resources. Omega on the other hand, follows a shared-state approach,
where multiple concurrent schedulers can view the whole cluster state, with
conflicts being resolved through a transactional mechanism~\cite{omega13}.
Tarcil leverages information on the type of resources applications need
to employ a sampling-base distributed scheduler that returns high quality resources
within a few milliseconds~\cite{Delimitrou15}.
Dejavu identifies a few workload classes and reuses previous
allocations for each class, to minimize reallocation
overheads~\cite{dejavu12}. CloudScale~\cite{Cloudscale},
PRESS~\cite{Gong10}, AGILE~\cite{agile} and the work by Gmach et
al.~\cite{Gmach07} predict future resource needs online, %perform online prediction of resource needs, often
often without a priori knowledge. Finally, auto-scaling systems,
such as Rightscale~\cite{Rightscale}, automatically scale the number of
physical or virtual instances used by webserving workloads,
to accommodate changes in user load.

A second line of work tries to identify resources that will allow a new, potentially-unknown application 
to meet its performance (throughput or tail latency) requirements~\cite{Delimitrou13, Delimitrou13d, Delimitrou13e, Delimitrou14, mars11a,Nathuji07,Mars13a}.
Paragon uses classification to determine the impact of platform heterogeneity
and workload interference on an unknown, incoming workload~\cite{Delimitrou13,Delimitrou13b}. It then
uses this information to achieve predictable performance, and high cluster utilization. % for the cluster.
Paragon, assumes that the cluster manager has full control over all
resources, which is often not the case in public clouds. Quasar extends the use of data mining 
in cluster management by additionally determining the appropriate amount of resources for a new 
application. Nathuji et al. developed a
feedback-based scheme that tunes resource assignments to mitigate
memory interference~\cite{Nathuji10}. Yang et al. developed an online scheme that
detects memory pressure and finds colocations that avoid interference
on latency-sensitive workloads~\cite{Mars13a}. Similarly, DeepDive
detects and manages interference between co-scheduled workloads in
a VM environment~\cite{Novakovic13}. 

Finally, CPI2~\cite{Zhang13} throttles
low-priority workloads that introduce destructive interference to important, latency-critical
services, using low-level metrics of performance collected through Google-Wide Profiling ({\smallcapital GWP}). 
In terms of managing platform heterogeneity, Nathuji et al.~\cite{Nathuji07}
and Mars et al.~\cite{Mars13b} quantified its impact on conventional
benchmarks and Google services, and designed schemes to predict the most
appropriate servers for a workload.

\vspace{0.05in}
\noindent{\bf{Cloud tracing \& diagnostics: }}
There is extensive related work on monitoring systems that has shown that execution traces can help diagnose performance, efficiency, and even security problems
in large-scale systems~\cite{xtrace,MysteryMachine,cloudseer,Xu16,dapper,GoogleTrace,Baek17,Rodrigues16,rootcause17}.
For example, X-Trace is a tracing framework that provides a comprehensive view of the behavior of services running 
on large-scale, potentially shared clusters. X-Trace supports several protocols and software systems, and has been deployed in several 
real-world scenarios, including DNS resolution, and a photo-hosting site~\cite{xtrace}. 
The Mystery Machine, on the other hand, leverages the massive amount of monitoring data cloud systems collect to determine the causal relationship between different requests~\cite{MysteryMachine}. 
Cloudseer serves a similar purpose, building an automaton for the workflow of each task based on normal execution, and then compares against this automaton at runtime to determine 
if the workflow has diverged from its expected behavior~\cite{cloudseer}. Finally, there are several production systems, including Dapper~\cite{dapper}, GWP~\cite{gwp}, and Zipkin~\cite{zipkin} which 
provide the tracing infrastructure for large-scale productions services at Google and Twitter, respectively. Dapper and Zipkin trace distributed user requests at RPC granularity, while GWP 
focuses on low-level hardware monitoring diagnostics. 

Root cause analysis of performance abnormalities in the cloud has also gained 
increased attention over the past few years, as the number of interactive, 
latency-critical services hosted in cloud systems has increased. 
Jayathilaka et al.~\cite{rootcause17}, for example, developed Roots, 
a system that automatically identifies the root cause of performance anomalies 
in web applications deployed in Platform-as-a-Service (PaaS) clouds. 
Roots tracks events within the PaaS cloud using a combination of metadata injection and platform-level instrumentation. 
Weng et al.~\cite{Weng17} similarly explore the cloud provider's ability 
to diagnose the root cause of performance abnormalities in multi-tier applications. 
Finally, Ouyang et al.~\cite{Ouyang16} focus 
on the root cause analysis of straggler tasks in distributed programming frameworks, 
like MapReduce and Spark. 

Even though this work does not specifically target interactive, latency-critical microservices, % and the diagnostics techniques are not learning-based,
or applications of similar granularity, such examples provide promising evidence 
that data-driven performance diagnostics can improve a large-scale system's ability 
to identify performance anomalies, and address them to meet its performance guarantees. %predictability and responsive

%\noindent{\bf{Root cause analysis: }}

\section{End-to-End Applications with Microservices}
\label{sec:methodology}

We motivate and evaluate Seer with a set of new end-to-end, interactive services built with microservices. Even though there are open-source 
microservices that can serve as components of a larger application, such as {\smallcapital\texttt{nginx}}~\cite{nginx}, {\smallcapital\texttt{memcached}}~\cite{memcached}, 
{\smallcapital\texttt{MongoDB}}~\cite{mongodb}, {\smallcapital\texttt{Xapian}}~\cite{Kasture16}, and %frameworks
{\smallcapital\texttt{RabbitMQ}}~\cite{rabbitmq}, there are currently no publicly-available end-to-end microservices applications, 
with the exception of a few simple architectures, like Go-microservices~\cite{gomicroservices}, and Sockshop~\cite{sockshop}. We design four end-to-end services implementing 
a \textit{Social Network}, a \textit{Media Service}, an \textit{E-commerce Site}, and a \textit{Banking System}. Starting from the Go-microservices architecture~\cite{gomicroservices}, we also
develop an end-to-end \textit{Hotel Reservation} system. Services are designed to be representative of frameworks used in production systems, modular, and easily reconfigurable. 
The end-to-end applications and tracing infrastructure are described in more detail and open-sourced in~\cite{Gan19}. %We are planning to open-source our applications and tracing infrastructure. 

Table~\ref{loc_stats} briefly shows the characteristics of each end-to-end application, 
including its communication protocol, the number of unique microservices it includes, and its 
breakdown by programming language and framework. 
Unless otherwise noted, all microservices are deployed in Docker containers. % to simplify deployment. % setup.
Below, we briefly describe the scope and functionality of each service. 
%; hand-written, and auto-generated by Thrift, where applicable. The majority of
%new code for the \textit{Social Network}, \textit{Media}, \textit{E-commerce}, and \textit{Banking} services goes towards the cross-microservice API, as well as a few microservices for which no open-source
%framework existed, e.g., assigning ratings to movies.
%with the remainder corresponding to a few simple new microservices.
%In the IoT application when compute happens at the edge, a large fraction of the code corresponds to new services, implemented to run on the specific drone hardware (ARM v7)~\cite{parrot},
%while when computation happens on the cloud we use readily available services for image recognition and obstacle avoidance, and most of the code goes towards
%cross-service APIs. % (this reflects the difference on Thrift-generated code as well).
%is aimed at
%connecting them with each other
%We also show the number of unique microservices for each application, and the breakdown per programming language.
%\vspace*{18\baselineskip}

\subsection{Social Network}

\noindent{\bf{Scope: }} The end-to-end service implements a broadcast-style social network with uni-directional follow relationships.

\vspace{0.06in}
\noindent{\bf{Functionality: }}Fig.~\ref{fig:social} shows the architecture of the end-to-end service. 
%The blue rectangles in each microservice signify tracing probe points. 
Users ({\smallcapital{\texttt{client}}}) %on the service
send requests over {\smallcapital\texttt{http}}, which first reach a load balancer, implemented with {\smallcapital\texttt{nginx}}, which selects a specific webserver is selected, also in
{\smallcapital\texttt{nginx}}. %the latter uses a \texttt{php-fpm} module to talk to the microservices responsible for composing and displaying posts, as well
%as microservices for advertisements, search engines, etc. All messages downstream of \texttt{php-fpm} are Apache Thrift RPCs~\cite{thrift}.
Users can create posts embedded with text, media, links, and tags to other users, which are then broadcasted to all their followers.
Users can also read, favorite, and repost posts, as well as reply publicly, or send a direct message to another user. The application also includes
machine learning plugins, such as ads and user recommender engines~\cite{Bottou,Netflix03,Witten,Kiwiel}, a search service using {\smallcapital\texttt{Xapian}}~\cite{Kasture16},
and microservices that allow users to follow, unfollow, or block other accounts. Inter-microservice messages use Apache Thrift RPCs~\cite{thrift}. 
The service's backend uses {\smallcapital\texttt{memcached}} for caching, and {\smallcapital\texttt{MongoDB}} for persistently storing posts, user profiles,
media, and user recommendations. 
%Finally, the service is instrumented with a distributed tracing system (Sec.~\ref{sec:tracing})
%to record RPC-level request latency.
This service is broadly deployed at Cornell and elsewhere, and currently has several hundred users. We use this installation 
to test the effectiveness and scalability of Seer in Section~\ref{sec:cloud_study}. 

\begin{figure}%{r}{0.73\textwidth}
\centering
\includegraphics[scale=0.352, viewport=40 0 700 440]{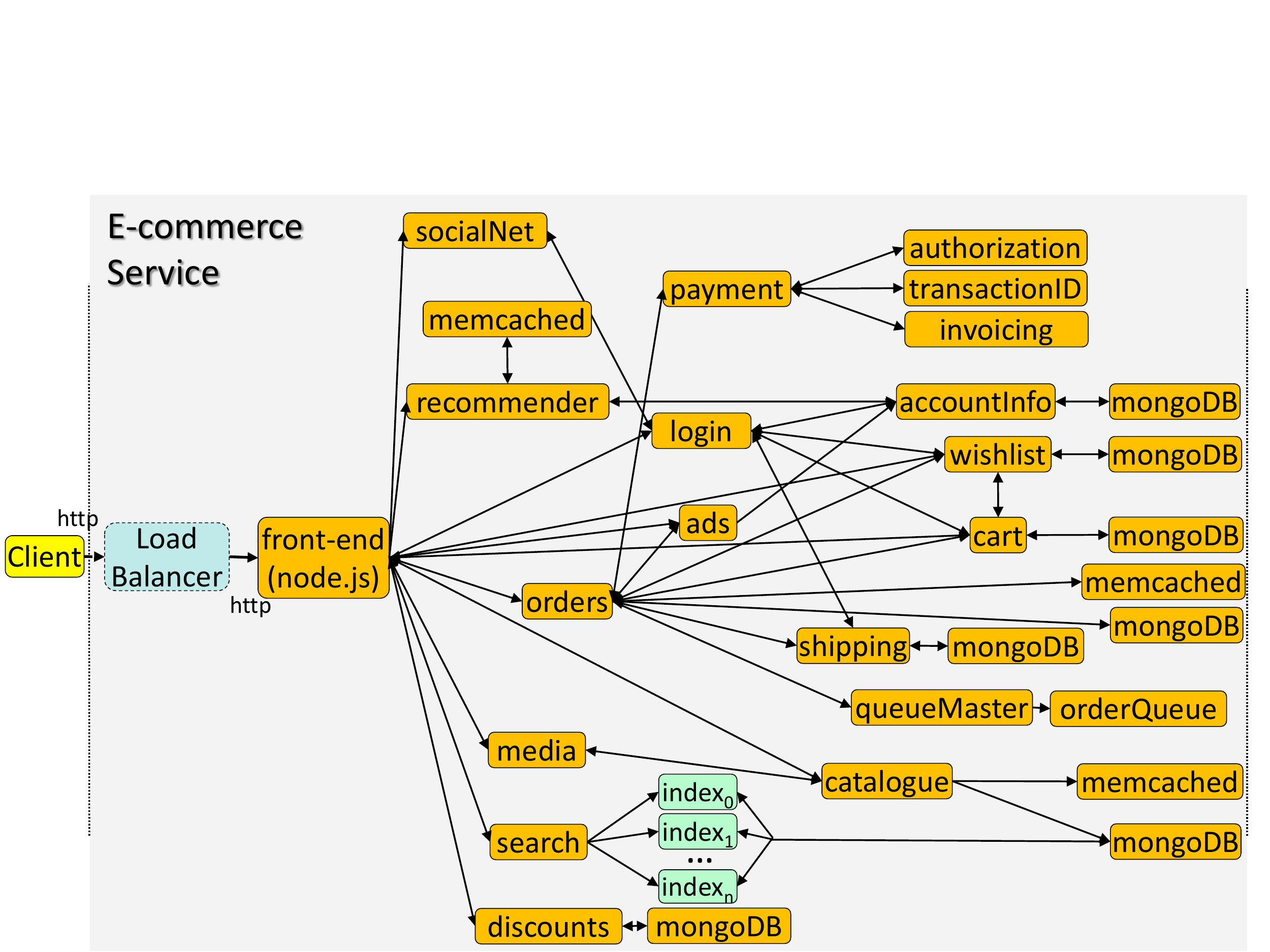}
\caption{\label{fig:ecommerce} {The architecture of the end-to-end \textit{E-commerce} application implementing an online clothing store.  }}
\end{figure}

\subsection{Media Service}

\noindent{\bf{Scope: }} The application implements an end-to-end service for browsing movie information, as well as reviewing, rating, renting, and streaming movies~\cite{Cockroft15,Cockroft16}.

\vspace{0.06in}
\noindent{\bf{Functionality: }} %Fig.~\ref{fig:movie} shows the architecture of the end-to-end service. %microservices graph used to load and display movie information. % a user sees when interacting with the service.
As with the social network, a client request hits the load balancer which distributes requests among multiple {\smallcapital\texttt{nginx}} webservers.
The front-end is similar to \textit{Social Network}, and users can search and browse information about movies, 
including the plot, photos, videos, and review information, as well as insert a review for a specific movie by logging in to their account.
Users can also select to rent a movie, which involves a payment authentication module to verify the user has enough funds, and
a video streaming module using {\smallcapital\texttt{nginx-hls}}, a production {\smallcapital\texttt{nginx}} module for HTTP live streaming.
Movie files are stored in NFS, to avoid the latency and complexity of accessing chunked records from non-relational databases,
while reviews are held in {\smallcapital\texttt{memcached}} and {\smallcapital\texttt{MongoDB}} instances. Movie information is maintained in a sharded and replicated MySQL DB.
%The application also includes services such as movie and advertisement recommenders, %and a search engine,
%as well as a couple auxiliary services for maintenance and service discovery, which are not shown in the figure.
We are similarly deploying \textit{Media Service} as a hosting site for project demos at Cornell, which students can browse and review. 
%widely at our institution; the former servers as an internal social network and currently has 582 registered users (165 active daily users on average), and the latter serves as a video hosting site
%for project demos that can be streamed and reviewed by the community. We use this Social Network deployment to study the tail at scale effects of microservices in Section~\ref{sec:tail_at_scale}

\subsection{E-Commerce Service}

\noindent{\bf{Scope: }}The service implements an e-commerce site for clothing. %, which can easily be generalized for other items.
The design draws inspiration, and uses several components of the open-source Sockshop application~\cite{sockshop}.

\vspace{0.06in}
\noindent{\bf{Functionality: }}The application front-end here is a {\smallcapital\texttt{node.js}} service.
Clients can use the service to browse the inventory using {\smallcapital\texttt{catalogue}}, a Go microservice that mines the back-end
{\smallcapital\texttt{memcached}} and {\smallcapital\texttt{MongoDB}} instances holding information about products. Users can also place {\smallcapital\texttt{orders}} (Go) 
by adding items to their {\smallcapital\texttt{cart}} (Java).
After they {\smallcapital\texttt{log in}} (Go) to their account, they can select {\smallcapital\texttt{shipping}} options (Java), process their {\smallcapital\texttt{payment}} (Go),
and obtain an {\smallcapital\texttt{invoice}} (Java) for their order. Orders are serialized and committed using {\smallcapital\texttt{QueueMaster}} (Go). Finally, the service includes
a {\smallcapital\texttt{recommender}} engine (C++), and microservices for creating {\smallcapital \texttt{wishlists}} (Java). %, and displaying current discounts.
%that interfaces with the \texttt{catalogue}, \texttt{user login} module,

\subsection{Banking System}

\noindent{\bf{Scope: }}The service implements a secure banking system, supporting payments, loans, and credit card management. %, or browse wealth management options.

\vspace{0.06in}
\noindent{\bf{Functionality: }}Users interface with a {\smallcapital\texttt{node.js}} front-end, similar to \textit{E-commerce}, to login to their account, search
information about the bank, or contact a representative. Once logged in, a user can process a payment from their account, pay their credit card or request a new one,
request a loan, and obtain information about wealth management options. Most microservices are written in Java and Javascript. The back-end databases
are {\smallcapital\texttt{memcached}} and {\smallcapital\texttt{MongoDB}} instances. The service also has a relational database ({\smallcapital\texttt{BankInfoDB}}) that includes information about
the bank, its services, and representatives. 

\subsection{Hotel Reservation Site}

\noindent{\bf{Scope: }}The service is an online hotel reservation site for browsing information about hotels, and making reservations. 

\vspace{0.06in}
\noindent{\bf{Functionality: }}The service is based on the Go-microservices open-source project~\cite{gomicroservices}, augmented with backend databases, and machine learning widgets for 
advertisement and hotel recommendations. A client request is first directed to one of the front-end webservers in node.js by a load balancer. The front-end then interfaces with 
the search engine, which allows users to explore hotel availability in a given region, and place a reservation. The service back-end consists of {\smallcapital\texttt{memcached}} %for in-memory caching, 
and {\smallcapital\texttt{MongoDB}} instances. %for persistent storage. 

%\subsection{Systems}

%First, we use a dedicated local cluster with 20, 2-socket 40-core servers with 128GB of RAM each. Each server is connected to a 40Gbps ToR switch over 10Gbe NICs. Second, 
%we deploy the Social Network service to Google Compute Engine (GCE) and Windows Azure clusters with hundreds of servers to study the scalability of Seer. 
%we use a 200-instance cluster on Google Compute Engine (GCE) to study the scalability of Seer. All instances are \texttt{n1-standard-64}, each with 64 vCPUs and 240GB of RAM.

\section{Seer Design} % Seer Techniques
\label{sec:design}

\begin{figure}
	\centering
	\includegraphics[scale=0.398, trim=0cm 0.2cm 0.4cm 8cm,clip=true]{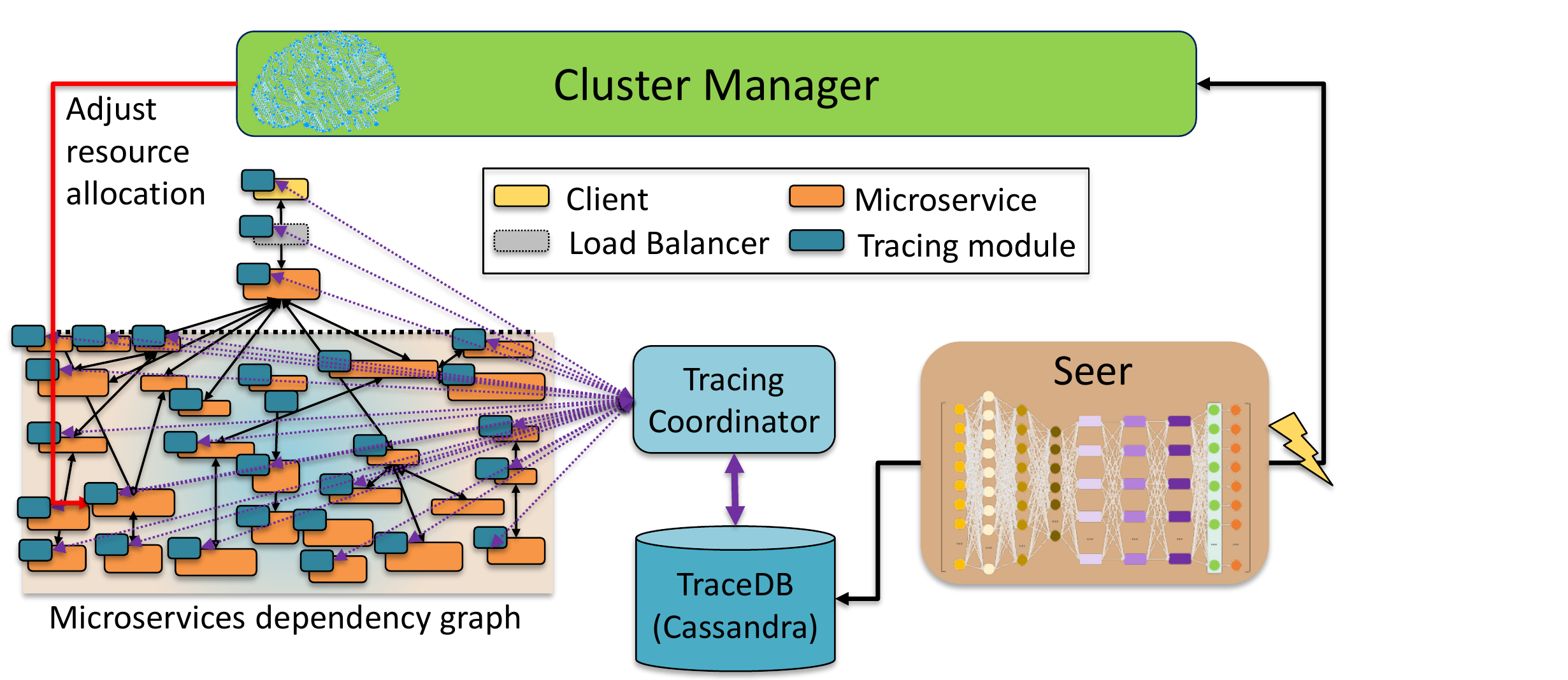}
	\caption{\label{fig:seer_overview} {Overview of \textit{Seer}'s operation. }}
\end{figure}

\subsection{Overview}
\label{sec:overview}

\begin{wrapfigure}[12]{l}{0.24\textwidth}
%\begin{figure}
	\vspace{-0.1in}
	\centering
	\includegraphics[scale=0.378, trim=0cm 0cm 4.4cm 9.8cm,clip=true]{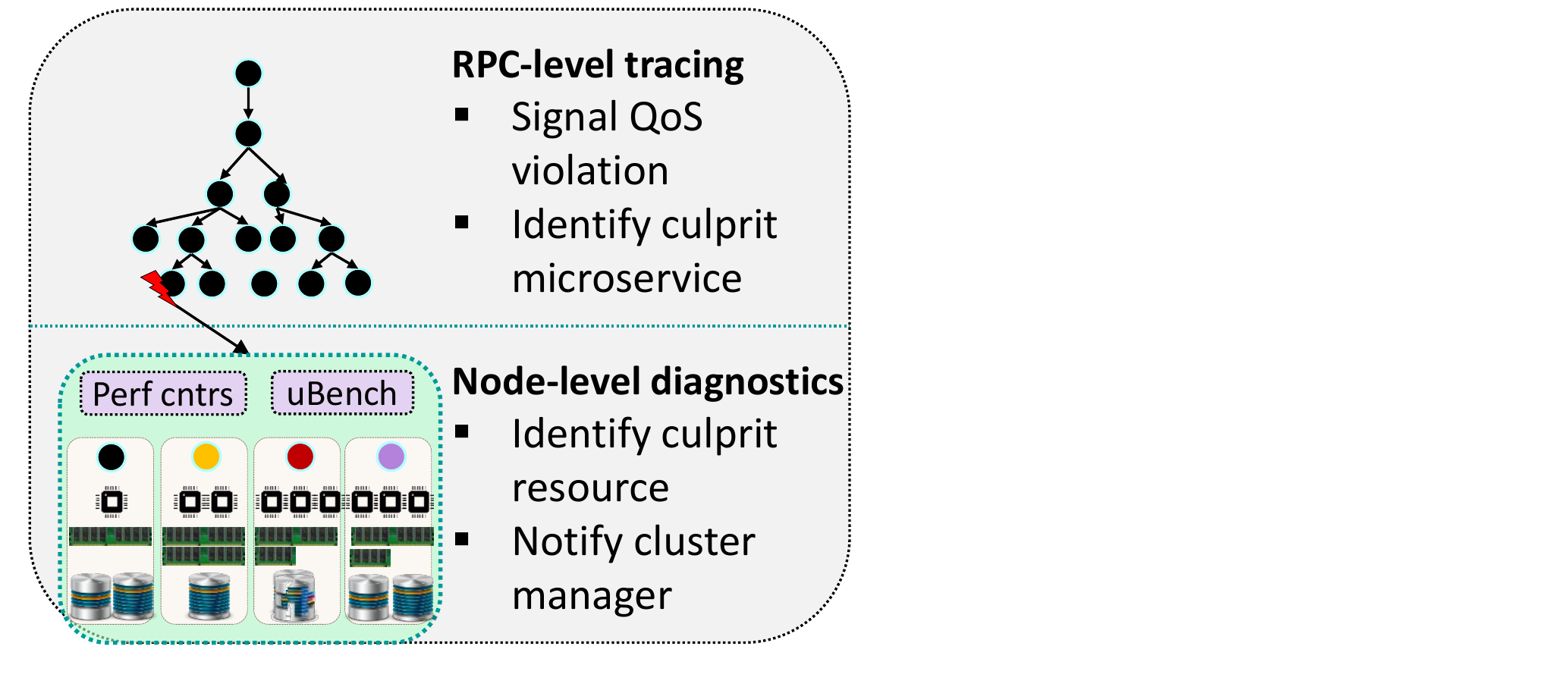}
	\vspace{-0.20in}
	\caption{\label{fig:seer_levels} {The two levels of tracing used in Seer.  }}
\end{wrapfigure}

Fig.~\ref{fig:seer_overview} shows the high-level architecture of the system. 
Seer is an online performance debugging system for cloud systems hosting interactive, latency-critical services. Even though we are focusing our analysis on microservices, 
where the impact of QoS violations is more severe, Seer is also applicable to general cloud services, and traditional multi-tier or Service-Oriented Architecture (SOA) workloads. 
Seer uses two levels of tracing, shown in Fig.~\ref{fig:seer_levels}. 

First, it uses a lightweight, distributed {\smallcapital RPC}-level tracing system, described in Sec.~\ref{sec:tracing}, which collects end-to-end execution 
traces for each user request, including per-tier latency and outstanding requests, associates {\smallcapital RPC}s belonging to the same end-to-end request, and aggregates them to a centralized Cassandra 
database (TraceDB). From there traces are used to train Seer to recognize patterns in space (between microservices) and time that 
lead to QoS violations. At runtime, Seer consumes real-time streaming traces to infer whether there is an imminent QoS violation. 

When a QoS violation is expected to occur and a culprit microservice has been located, Seer uses its lower tracing level, 
which consists of detailed per-node, low-level hardware monitoring primitives, such as performance counters, 
to identify the reason behind the QoS violation. It also uses this information to provide the cluster manager with 
recommendations on how to avoid the performance degradation altogether. When Seer runs on a public cloud where 
performance counters are disabled, it uses a set of tunable microbenchmarks to determine the source of unpredictable performance (see Sec.~\ref{sec:per_node}). 
Using two specialized tracing levels instead of collecting detailed low-level traces for all active microservices 
ensures that the distributed tracing is lightweight enough to track all active requests and services in the system, and that detailed 
low-level hardware tracing is only used on-demand, for microservices likely to cause performance disruptions. 
%For Seer to be effective in improving performance predictability, inference needs to occur with enough slack for the cluster manager's 
%action to take effect. 
In the following sections we describe the design of the tracing system, the learning techniques in Seer and its 
low-level diagnostic framework, and the system insights we can draw from Seer's decisions to improve cloud application design. 
%multi-tier applications, and microservices. description of Fig. 8
%The distributed tracing system described in Sec.~\ref{sec:tracing} collects end-to-end execution traces 
%for each user request, including per-tier latency and outstanding requests, and aggregates them to a centralized Cassandra database (TraceDB). 

%If one is expected, Seer first uses its per-node notifies 

%main parts of system, process at a high level

\subsection{Distributed Tracing}
\label{sec:tracing}

\begin{figure}
	\centering
	\includegraphics[scale=0.48, trim=-0.5cm 0 0.6cm 3.8cm, clip=true]{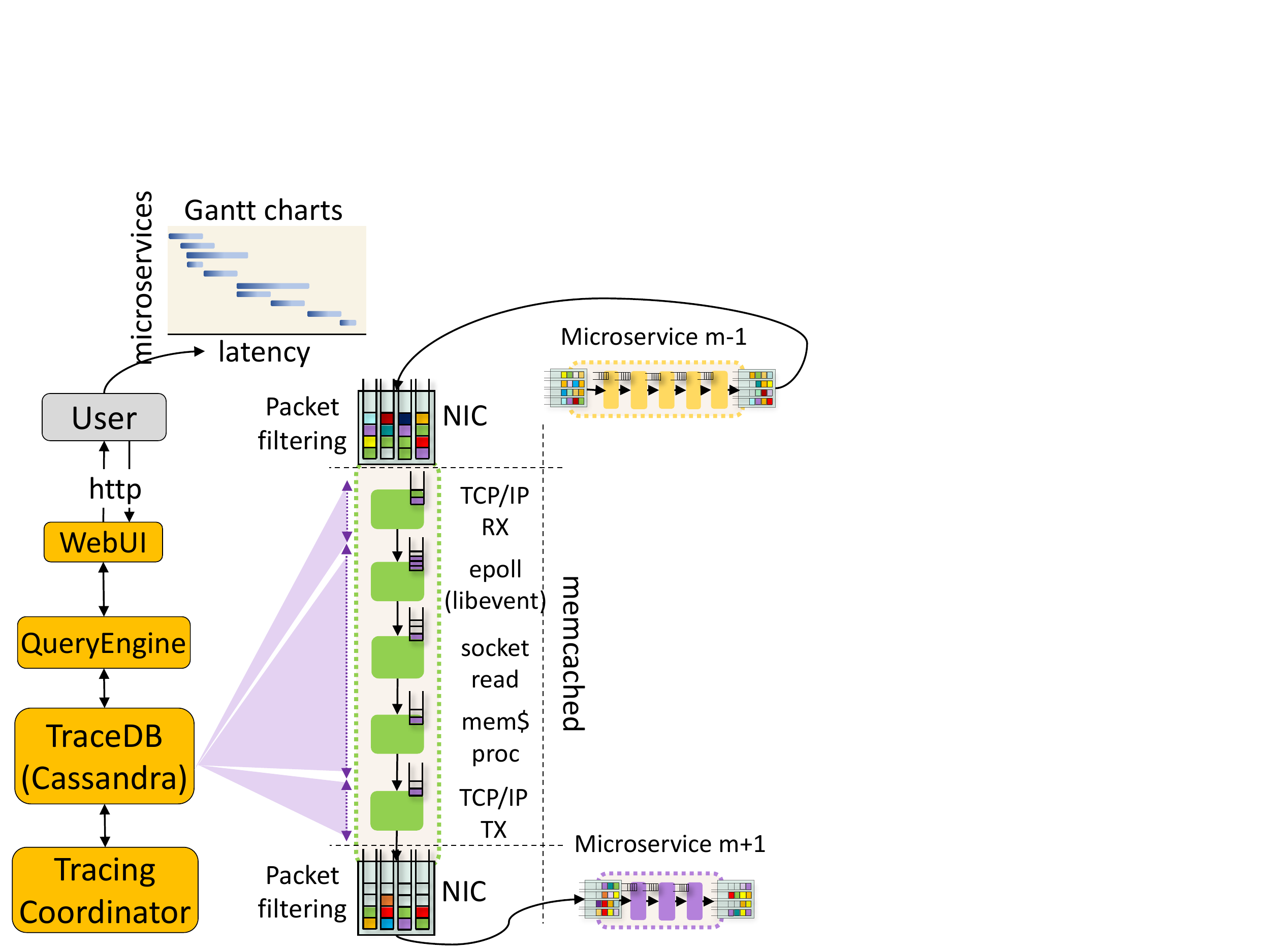}
	\caption{\label{fig:seer_instrumentation} {Distributed tracing and instrumentation in Seer. }}
\end{figure}

A major challenge with microservices is that one cannot simply rely on the client to report performance, as with traditional client-server applications.
We have developed a distributed tracing system for Seer, similar in design to Dapper~\cite{dapper} and Zipkin~\cite{zipkin} 
that records per-microservice latencies, % upon an RPC's arrival and departure, 
using the Thrift timing interface, as shown in Fig.~\ref{fig:seer_instrumentation}. %RPCs and REST requests are timestamped upon arrival and departure from each microservice by the tracing module,
%The design of the tracing system is similar to Dapper~\cite{dapper} and Zipkin~\cite{zipkin}. 
%accumulated by the \texttt{Trace Collector}, implemented similarly
%to the Zipkin Collector~\cite{zipkin}, and stored in a centralized Cassandra database. 
We additionally track the number of requests queued in each microservice (\textit{outstanding requests}), since queue lengths 
are highly correlated with performance and QoS violations~\cite{ubik,queueing,nsdi18,Yu11}.  
In all cases, the overhead from tracing without request sampling is negligible, less than $0.1\%$ on end-to-end latency, and less than $0.15\%$ on throughput (QPS), which is tolerable for such systems~\cite{dapper,MysteryMachine,gwp}.
Traces from all microservices are aggregated in a centralized database~\cite{cassandra}. 

%\begin{figure}
%	\centering
%	\begin{tabular}{cc}
%		\includegraphics[scale=0.36, trim=0.2cm 0.4cm 13.2cm 3.2cm, clip=true]{figures/tracing7.pdf} & 
%	\includegraphics[scale=0.34, trim=3.4cm -1.2cm 0cm 6.2cm,clip=true]{figures/clock_sync3.pdf}
%\end{tabular}
%\caption{\label{fig:tracing} (a) The distributed RPC-level tracing system and (b) its clock synchronization mechanism. }
%\end{figure}

%- explain tracing, how you get the latency per microservice (rely on nginx), how you distinguish between network and app processing time

\vspace{0.06in}
\noindent{\bf{Instrumentation: }}The tracing system requires two types of application instrumentation. 
First, to distinguish between the time spent 
processing network requests, and the time that goes towards application computation, 
we instrument the application to report the time it sees a new request (post-{\smallcapital RPC} processing). 
We similarly instrument the transmit side of the application. Second, systems have multiple sources of queueing 
in both hardware and software. To obtain accurate measurements of queue lengths per microservice, we need 
to account for these different queues. Fig.~\ref{fig:seer_instrumentation} shows an example of Seer's 
instrumentation for memcached. Memcached includes five main stages~\cite{Leverich14}, 
{\smallcapital TCP/IP} receive, \texttt{epoll}/\texttt{libevent}, 
reading the request from the socket, processing the request, and responding over {\smallcapital TCP/IP}, 
either with the \texttt{<k,v>} pair for a read, or with an \texttt{ack} for a write. 
Each of these stages includes a hardware ({\smallcapital NIC}) or software (\texttt{epoll},\texttt{socket read},\texttt{memcached proc}) queue. 
For the {\smallcapital NIC} queues, Seer filters packets based on the destination microservice, but accounts for 
the aggregate queue length if hardware queues are shared, since that will impact how fast a microservice's packets get processed. 
For the software queues, Seer inserts probes in the application to read the number of queued requests in each case. %length of each of these queues. 

%and additionally distinguish between the time spent processing network requests and the time that goes towards application computation.
\vspace{0.06in}
\noindent{\bf{Limited instrumentation: }}As seen above, accounting for all sources of queueing in a complex system requires non-trivial instrumentation. 
This can become cumbersome if users leverage third-party applications in their services, 
or in the case of public cloud providers which do not have access to the source code 
of externally-submitted applications for instrumentation. 
%reuse open-source microservices which make instrumenting the source code cumbersome. Similarly, cloud providers often have no visibility 
%on the source code of externally-submitted applications. %, which would preclude using a system like Seer. 
%Third party apps... 
In these cases Seer relies on the requests queued exclusively in the {\smallcapital NIC} to signal upcoming QoS violations. 
In Section~\ref{sec:validation} we compare the accuracy of the full versus limited instrumentation, and see that using network queue depths alone is enough 
to signal a large fraction of QoS violations, although smaller than when the full instrumentation is available. Exclusively polling 
{\smallcapital NIC} queues identifies hotspots caused by routing, incast, failures, and resource saturation, 
but misses QoS violations that are caused by performance and 
efficiency bugs in the application implementation, such as blocking behavior between microservices. 
Signaling such bugs helps developers better understand the microservices model, 
and results in better application design. 

%Given the accuracy of Seer when only using queue depths in the NIC, it seems a disproportionate amount of effort to instrument 
%microservices for a XXX\% increase in detection accuracy. However, QoS violations caused by queues within the application often reveal performance and efficiency bugs, such as blocking behavior between 
%application functions, which help users improve the design of microservices-based applications to begin with. 

\vspace{0.06in}
\noindent{\bf{Inferring queue lengths: }}
Additionally, there has been recent work on using deep learning to reverse engineer the number of queued requests 
in switches across a large network topology~\cite{nsdi18}, when tracing 
information is incomplete. Such techniques are also applicable and beneficial for Seer when the default level of 
instrumentation is not available. %if you're missing some traces at some point, not without any tracing at all. 

%\noindent{\bf{Trace synchronization: }}Because different microservices of a single end-to-end service run on different physical machines, there is a chance that 
%the clocks of the different servers fall out of sync~\cite{nsdi18}. This can impact Seer's ability to correlate consecutive events such as queue build ups, or QoS violations. 
%We periodically synchronize server clocks by sending a few small RPC requests 

\begin{table}
	\skiourakisize
	\begin{tabular}{cp{0.4cm}p{0.27cm}p{0.3cm}p{0.3cm}p{0.3cm}ccp{0.4cm}}
	\hline
	\multirow{2}{*}{Name} & \multirow{2}{*}{LoC} & \multicolumn{4}{c}{Layers} & {Nonlinear} & \multirow{2}{*}{Weights} & {Batch} \\
	\cline{3-6}
	& & FC & \multicolumn{1}{c}{Conv} & Vect & Total & Function & & Size \\
	\hline
	CNN & 1456 & & 8 & & 8 & ReLU & 30K & 4 \\
	\hdashline[0.5pt/2.5pt]
	LSTM & 944 & 12 &  & 6 & 18 & sigmoid,tanh & 52K & 32 \\
	\hdashline[0.5pt/2.5pt]
	\multirow{2}{*}{Seer} & \multirow{2}{*}{2882} & \multirow{2}{*}{10} & \multirow{2}{*}{7} & \multirow{2}{*}{5} & \multirow{2}{*}{22} & ReLU & \multirow{2}{*}{80K} & \multirow{2}{*}{32} \\
	 & & & & & & sigmoid,tanh & & \\
	\hline
\end{tabular}
\caption{\label{NNs} The different neural network configurations we explored for Seer. }
\end{table}

%\begin{figure}
%	\centering
%	\includegraphics[scale=0.30, trim=0cm 0cm 0.4cm 10cm,clip=true]{figures/backpressure.pdf}
%	\caption{\label{fig:backpressure} {Backpressure.  }}
%\end{figure}

\subsection{Deep Learning in Performance Debugging}

A popular way to model performance in cloud systems, especially when there are dependencies between tasks, are %model-based 
%approaches like 
\textit{queueing networks}~\cite{queueing}. Although queueing networks are a valuable tool to 
model how bottlenecks propagate through the system, they require in-depth knowledge of application semantics and structure, and can become overly complex as 
applications and systems scale. They additionally cannot easily capture all sources of contention, such as the {\smallcapital OS} and network stack. 
%they do not capture all sources of contention in real systems, including the operating system and the network stack, 
%and are therefore not accurate enough to predict upcoming QoS violations. 

%\begin{wrapfigure}[16]{l}{0.24\textwidth}
%	\centering
%	\includegraphics[scale=0.41, trim=3cm 1.8cm 0cm 5.4cm,clip=true]{figures/clock_sync3.pdf}
%	\caption{\label{fig:seer_overview} {Clock synchronization mechanism. }}
%\end{wrapfigure}

Instead in Seer, we take a data-driven, application-agnostic approach that assumes no information 
about the structure and bottlenecks of a service, making it robust to unknown and changing applications, 
and relying instead on practical learning techniques to infer patterns that lead to QoS violations. %learning instead patterns making the system robust to changing and unknown services. 
This includes both \textit{spatial} patterns, such as dependencies between microservices, 
and \textit{temporal} patterns, such as input load, and resource contention. 
The key idea in Seer is that conditions that led to QoS violations in the past 
can be used to anticipate unpredictable performance in the near future. 
Seer uses execution traces annotated with QoS violations and collected over time 
to train a deep neural network to signal upcoming QoS violations. 
Below we describe the structure of the neural network, why deep learning is 
well-suited for this problem, and how Seer adapts to changes in application structure online.

\vspace{0.06in}
\noindent{\bf{Using deep learning: }}Although deep learning is not the only approach that can be used 
for proactive QoS violation detection, there are several reasons why it is preferable in this case. 
First, the problem Seer must solve is a pattern matching problem of recognizing conditions that result 
in QoS violations, where the patterns are not always known in advance or easy to annotate. 
This is a more complicated task than simply signaling a microservice with many enqueued requests, 
for which simpler classification, regression, or sequence labeling techniques would suffice~\cite{Netflix03,Bottou,Witten}. 
Second, the {\smallcapital DNN} in Seer assumes no a priori knowledge about dependencies 
between individual microservices, making it applicable to frequently-updated services, where describing 
changes is cumbersome or even difficult for the user to know. 
%services where the application architecture changes frequently, is overly complex for users to manually express dependencies, or for public cloud settings where the cloud provider does not 
%have access to the application source code. 
Third, deep learning has been shown to be especially effective in pattern recognition problems with massive datasets, %where datasets are plentiful, %where the pattern is unknown or too complex for users to annotate, 
e.g., in image or text recognition~\cite{tensorflow}. 
Finally, as we show in the validation section (Sec.~\ref{sec:validation}), deep learning allows Seer to recognize QoS 
violations with high accuracy in practice, and within the opportunity window 
the cluster manager has to apply corrective actions. %We are currently exploring more scalable deep learning approaches that are robust to... changes?? 

\begin{figure}
	\centering
	\includegraphics[scale=0.34, viewport = 55 0 635 420]{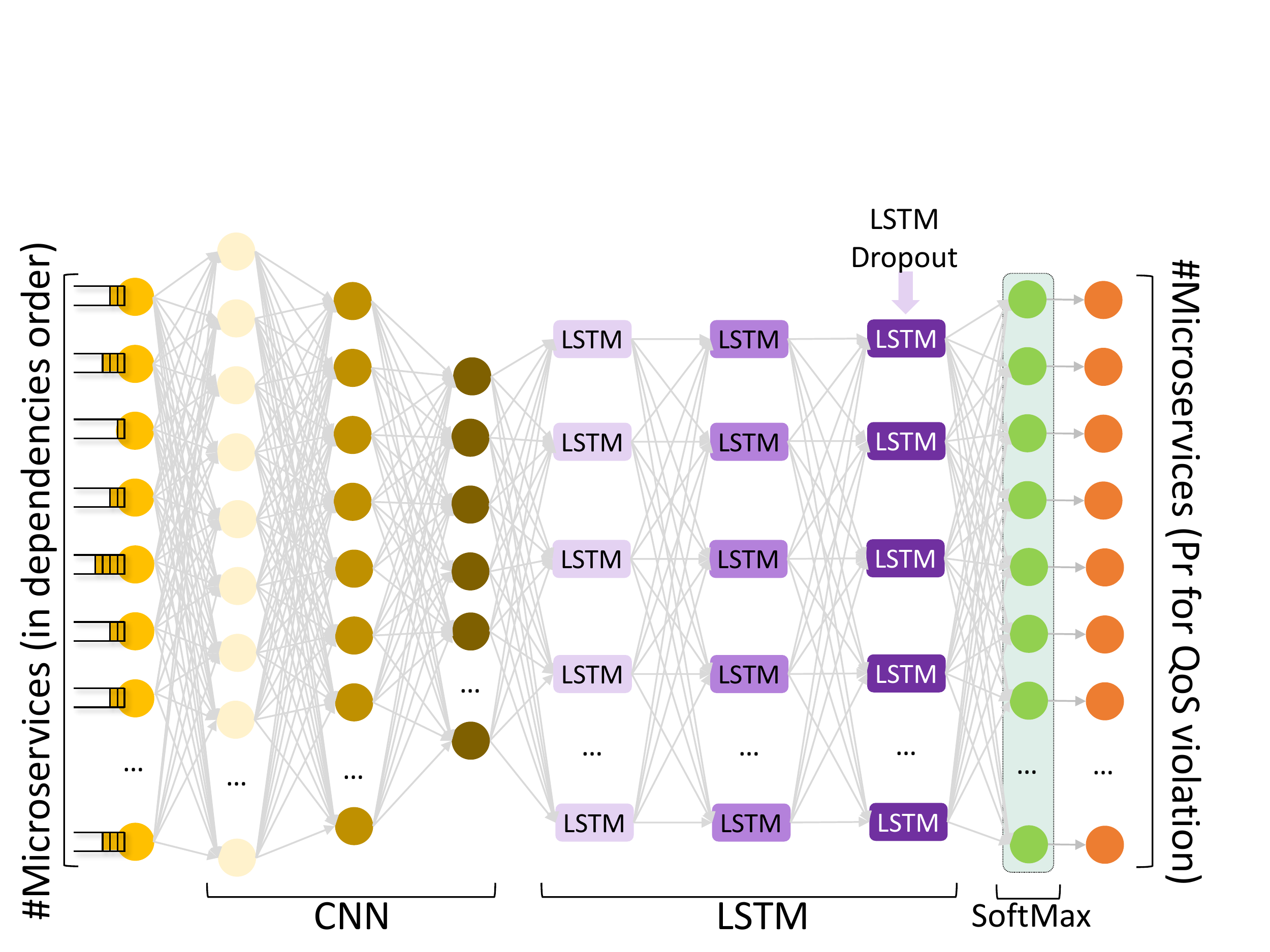}
	\caption{\label{fig:seer} The deep neural network design in Seer, consisting of a set of convolution layers followed by a set of long-short term memory layers. Each input and output neuron corresponds to a 
	microservice, ordered in topological order, from back-end microservices in the top, to front-end microservices in the bottom. }
\end{figure}

\vspace{0.06in}
%\noindent{\bf{Selecting a DNN architecture: }}
\noindent{\bf{Configuring the DNN: }}The input used in the network is essential for its accuracy. 
We have experimented with resource utilization, latency, and queue depths as input metrics. 
Consistent with prior work, utilization is not a good proxy for performance~\cite{Lo14,ubik,Delimitrou13,Delimitrou16}. 
Latency similarly leads to many false positives, or to incorrectly pinpointing computationally-intensive microservices as QoS violation culprits. 
Again consistent with queueing theory~\cite{queueing} and prior work~\cite{nsdi18,ubik,Delimitrou15,Delimitrou14,amdahls18}, 
per-microservice queue depths accurately capture performance bottlenecks and pinpoint the microservices causing them. 
We compare against utilization-based approaches in Section~\ref{sec:validation}. 
The number of input and output neurons is equal to the number of active microservices in the cluster, 
with the input value corresponding to queue depths, and the output value 
to the probability for a given microservice to initiate a QoS violation. 
Input neurons are ordered according to the topological application structure, 
with dependent microservices corresponding to consecutive neurons to capture 
hotspot patterns in space. Input traces are annotated with the microservices 
that caused QoS violations in the past. 

The choice of {\smallcapital DNN} architecture is also instrumental to its accuracy. 
There are three main {\smallcapital DNN} designs that are popular today: fully connected networks ({\smallcapital FC}), 
convolutional neural networks ({\smallcapital CNN}), and recurrent neural networks ({\smallcapital RNN}), 
especially their Long Short-Term Memory ({\smallcapital LSTM}) class. 
For Seer to be effective in improving performance predictability, 
inference needs to occur with enough slack for the cluster manager's action to take effect. 
Hence, we focus on the more computationally-efficient {\smallcapital CNN} and {\smallcapital LSTM} networks. 
{\smallcapital CNN}s are especially effective at reducing the dimensionality of large datasets, 
and finding patterns in space, e.g., in image recognition. 
{\smallcapital LSTM}s, on the other hand, are particularly effective at finding patterns in time, 
e.g., predicting tomorrow's weather based on today's measurements. 
Signaling QoS violations in a large cluster requires both spatial recognition, 
namely identifying problematic clusters of microservices whose dependencies cause QoS violations and discarding noisy but non-critical microservices, 
and temporal recognition, namely using past QoS violations to anticipate future ones. 
We compare three network designs, a {\smallcapital CNN}, a {\smallcapital LSTM}, and 
a hybrid network that combines the two, using the {\smallcapital CNN} first to reduce 
the dimensionality and filter out microservices that do not affect 
end-to-end performance, and then an {\smallcapital LSTM} with a \textit{SoftMax} final layer 
to infer the probability for each microservice to initiate a QoS violation. 
The architecture of the hybrid network is shown in Fig.~\ref{fig:seer}. 
Each network is configured using hyperparameter tuning to avoid overfitting, 
and the final parameters are shown in Table~\ref{NNs}. 

We train each network on a week's worth of trace data collected on a 20-server cluster 
running all end-to-end services (for methodology details see Sec.~\ref{sec:validation})
%, and its parameters (number of hidden layers, learning rate $a$, regularization, batch size, and units per hidden layer) are tuned to avoid overfitting. 
and test it on traces collected on a different week, after the servers had been patched, 
and the OS had been upgraded. 

\begin{wrapfigure}[13]{r}{0.24\textwidth}
	\centering
		\includegraphics[scale=0.25, viewport = 45 0 455 380]{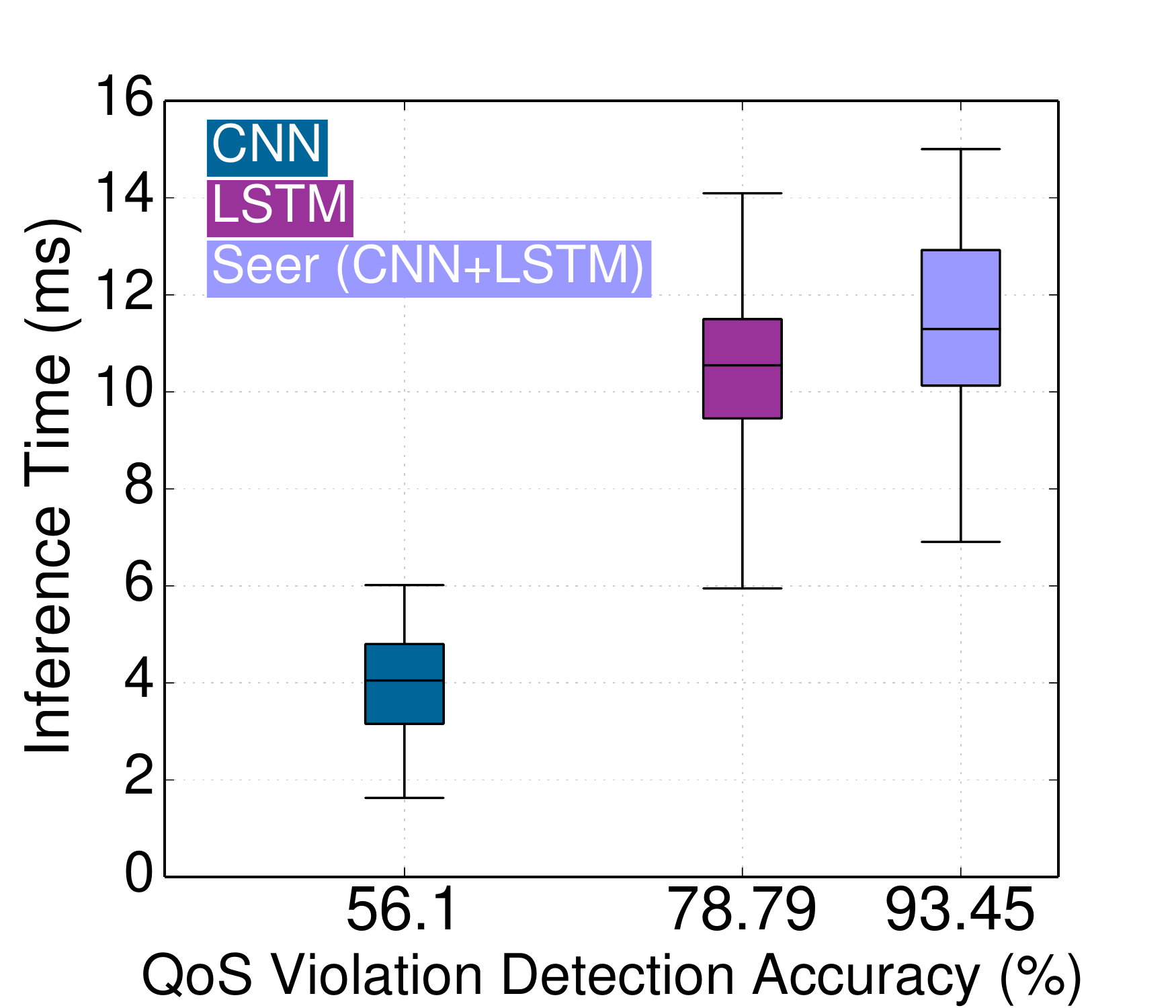}
	\caption{\label{fig:comparison} Comparison of DNN architectures. }
\end{wrapfigure}

The quantitative comparison of the three networks is shown in Fig.~\ref{fig:comparison}. 
The {\smallcapital CNN} is by far the fastest, but also the worst performing, 
since it is not designed to recognize patterns in time that lead to QoS violations. The {\smallcapital LSTM} on the 
other hand is especially effective at capturing load patterns over time, 
but is less effective at reducing the dimensionality of the original dataset, 
which makes it prone to false positives due to microservices with many outstanding requests, 
which are off the critical path. It also incurs higher overheads for inference 
than the {\smallcapital CNN}. Finally, Seer correctly anticipates 93.45\% of violations, 
outperforming both networks, for a small increase in inference time compared to {\smallcapital LSTM}. 
Given that most resource partitioning decisions take effect after a few 100ms, 
the inference time for Seer is within the window of opportunity the cluster manager has to take action. 
More importantly it attributes the QoS violation to the correct microservice, 
simplifying the cluster manager's task. % of taking action to avoid it. 
QoS violations missed by Seer included four random load spikes, % in user load, 
and a network switch failure which caused high packet drops. 

Out of the five end-to-end services, the one most prone to QoS violations initially was \textit{Social Network}, 
first, because it has stricter QoS constraints than e.g., \textit{E-commerce}, and second, 
due to a synchronous and cyclic communication between three neighboring services that caused them 
to enter a positive feedback loop until saturation. We reimplemented the communication protocol between them post-detection. 
On the other hand, the service for which QoS violations were hardest to detect was \textit{Media Service}, 
because of a faulty memory bandwidth partitioning mechanism in one of our servers, 
which resulted in widely inconsistent memory bandwidth allocations during movie streaming. 
Since the QoS violation only occurred when the specific streaming microservice was scheduled 
on the faulty node, it was hard for Seer to collect enough datapoints to signal the violation. 

\vspace{0.06in}
\noindent{\bf{Retraining Seer: }}%Microservices are designed to allow frequent updates. 
%While this is necessary to improve application design, it requires partially retraining Seer's network. 
By default training happens once, and can be time consuming, taking several hours up to a day for week-long traces 
collected on our 20-server cluster (Sec.~\ref{sec:validation} includes a detailed sensitivity study for training time). 
However, one of the main advantages of microservices is that they simplify frequent application updates, with 
old microservices often swapped out and replaced by newer modules, or large services progressively broken down to microservices. 
If the application (or underlying hardware) change significantly, Seer's detection accuracy can be impacted. 
To adjust to changes in the execution environment, Seer retrains incrementally in the background, 
using the \textit{transfer learning}-based approach in~\cite{syed17}. Weights from previous training rounds are stored in disk, 
allowing the model to continue training from where it last left off when new data arrive, reducing the training time by 2-3 orders of magnitude. 
Even though this approach allows Seer to handle application changes almost in real-time, 
it is not a long-term solution, since new weights are still polluted by the previous application architecture. 
When the application changes in a major way, e.g., microservices on the critical path change, 
Seer also retrains from scratch in the background. While the new network trains, QoS violation 
detection happens with the incrementally-trained interim model. 
In Section~\ref{sec:validation}, we evaluate Seer's ability 
to adjust its estimations to application changes. % in application design. 

\subsection{Hardware Monitoring}
\label{sec:per_node}

Once a QoS violation is signaled and a culprit microservice is pinpointed, Seer uses low-level monitoring to identify 
the reason behind the QoS violation. The exact process depends on whether Seer has access to performance counters. %low-level monitoring probes. 

\vspace{0.03in}
\noindent{\bf{Private cluster: }}When Seer has access to hardware events, such as performance counters, it uses them to determine 
the utilization of different shared resources. Note that even though utilization is a problematic metric for anticipating QoS violations
in a large-scale service, once a culprit microservice has been identified, examining the utilization of different resources can provide 
useful hints to the cluster manager on suitable decisions to avoid degraded performance. 
Seer specifically examines {\smallcapital CPU}, memory capacity and bandwidth, network bandwidth, cache contention, 
and storage I/O bandwidth when prioritizing a resource to adjust. 
Once the saturated resource is identified, Seer notifies the cluster manager to take action. 

\vspace{0.03in}
\noindent{\bf{Public cluster: }}When Seer does not have access to performance counters, it instead uses a set of 10 tunable contentious microbenchmarks, 
each of them targeting a different shared resource~\cite{Delimitrou13b} to determine resource saturation. For example, if Seer injects the memory bandwidth 
microbenchmark in the system, and tunes up its intensity without an impact on the co-scheduled microservice's performance, memory bandwidth is most likely 
not the resource that needs to be adjusted. Seer starts from microbenchmarks corresponding to core resources, and progressively moves to resources further away 
from the core, until it sees a substantial change in performance when running the microbenchmark. Each microbenchmark takes approximately 10ms to complete, avoiding 
prolonged degraded performance. 

Upon identifying the problematic resource(s), Seer notifies the cluster manager, 
which takes one of several resource allocation actions, %action via one of several resource allocation mechanisms, 
resizing the Docker container, or using mechanisms like 
Intel's Cache Allocation Technology ({\smallcapital CAT}) for last level cache ({\smallcapital LLC})
partitioning, and the Linux traffic control's hierarchical token bucket ({\smallcapital HTB}) 
queueing discipline in \texttt{qdisc}~\cite{qdisc,Lo15} for network bandwidth partitioning.

%high utilization is neither a necessary nor a sufficient condition for 
%a QoS violation, which resource is saturated

%- post-detection diagnostics

%- using low-level hardware monitoring and microbenchmarks to detect root cause of QoS violation

%- training \& inference

%\subsection{QoS Violation Prevention}

\subsection{System Insights from Seer}
\label{sec:insights}

Using learning-based, data-driven approaches in systems is most useful when these techniques are used to gain insight into system problems, instead 
of treating them as black boxes. Section~\ref{sec:validation} 
includes an analysis of the causes behind QoS violations signaled by Seer, 
including application bugs, poor resource provisioning decisions, and hardware failures. 
Furthermore, we have deployed Seer in a large installation of 
the \textit{Social Network} service over the past few months, 
and its output has been instrumental not only in guaranteeing QoS, but in understanding 
sources of unpredictable performance, and improving the application design. % accordingly. 
This has resulted both in progressively fewer QoS violations over 
time, and a better understanding of the design challenges of microservices. %experience for the end users. 

%- show implementation faults, bad communication patterns, use ML not as a black box, but to gain insight into complex systems... 

%Later show that errors are decreasing

\subsection{Implementation}

Seer is implemented in 12{\smallcapital KLOC} of C,C++, and Python. 
It runs on Linux and OSX and supports applications in various languages, 
including all frameworks the end-to-end services are designed in. 
Furthermore, we provide automated patches for the instrumentation probes 
for many popular microservices, including {\smallcapital\texttt{NGINX}}, 
{\smallcapital\texttt{memcached}}, {\smallcapital\texttt{MongoDB}}, 
{\smallcapital\texttt{Xapian}}, and all Sockshop and \textit{Go-microservices} 
applications to minimize the development effort from the user's perspective. 

Seer is a centralized system; we use master-slave mirroring to improve fault tolerance, 
with two hot stand-by masters that can take over if the primary system fails. 
Similarly, the trace database is also replicated in the background. 

\noindent{\bf{Security concerns: }}Trace data is stored and processed unencrypted in Cassandra. 
Previous work has shown that the sensitivity applications have to different resources can leak information 
about their nature and characteristics, making them vulnerable to malicious security attacks~\cite{Delimitrou17,Zhao18,Huang14,Xu11,Gupta13b,Darwish13,Varadarajan14,Varadarajan15,Shue12}. 
Similar attacks are possible using the data and techniques in Seer, and are deferred to future work. %. For the scope of this paper, we assume that Seer is not malicious; we defer such considerations to future work.  

\section{Seer Analysis and Validation}
\label{sec:validation}

\subsection{Methodology}

\noindent{\bf{Server clusters: }}First, we use a dedicated local cluster with 20, 2-socket 40-core servers with 128GB of RAM each. Each server is connected to a 40Gbps ToR switch over 10Gbe NICs. Second,
we deploy the Social Network service to Google Compute Engine (GCE) and Windows Azure clusters with hundreds of servers to study the scalability of Seer.

\noindent{\bf{Applications: }}We use all five end-to-end services of Table~\ref{loc_stats}. Services for now are driven by open-loop workload generators, and the input load varies 
from constant, to diurnal, to load with spikes in user demand. In Section~\ref{sec:cloud_study} 
we study Seer in a real large-scale deployment of the \textit{Social Network}; in that case the input load 
is driven by real user traffic. 

%- input load for applications. 

%\begin{figure}
%\centering
%\includegraphics[scale=0.218, viewport = 35 -70 455 322]{figures/False.pdf}
%\caption{\label{fig:false}
%The false negatives and false positives in Seer as we vary the prediction window. }
%\end{figure}

\subsection{Evaluation}

%\noindent{\bf{Inference time: }}As seen in Fig.~\ref{fig:comparison}, inference time for Seer varies between 6.8ms and 14.8ms. This is enough for the cluster manager to apply corrective action in all cases. 
\begin{figure}
	\centering
	\begin{tabular}{cc}
		\includegraphics[scale=0.26, viewport = 55 50 400 410]{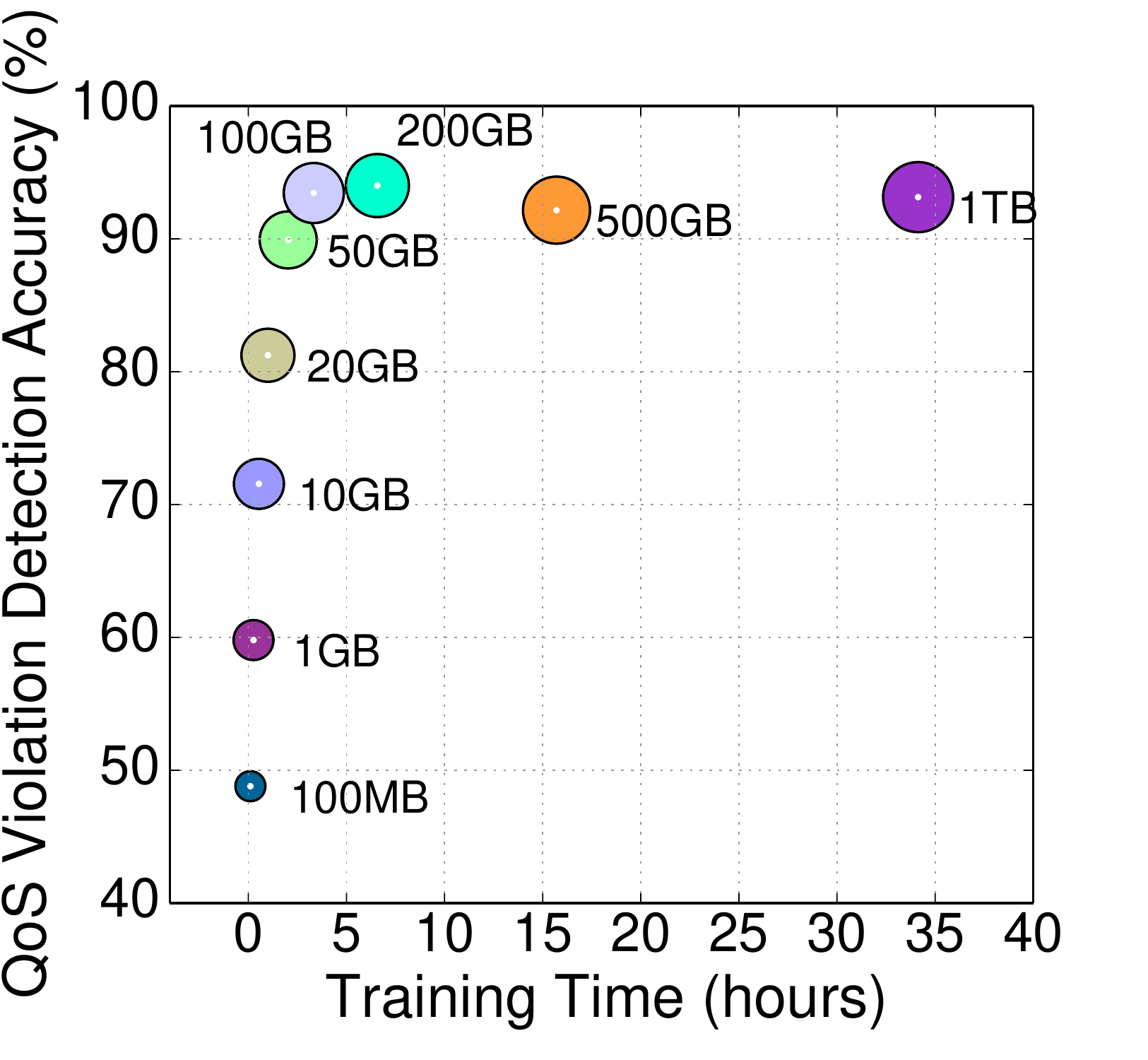} &
		\includegraphics[scale=0.26, viewport = 5 50 455 410]{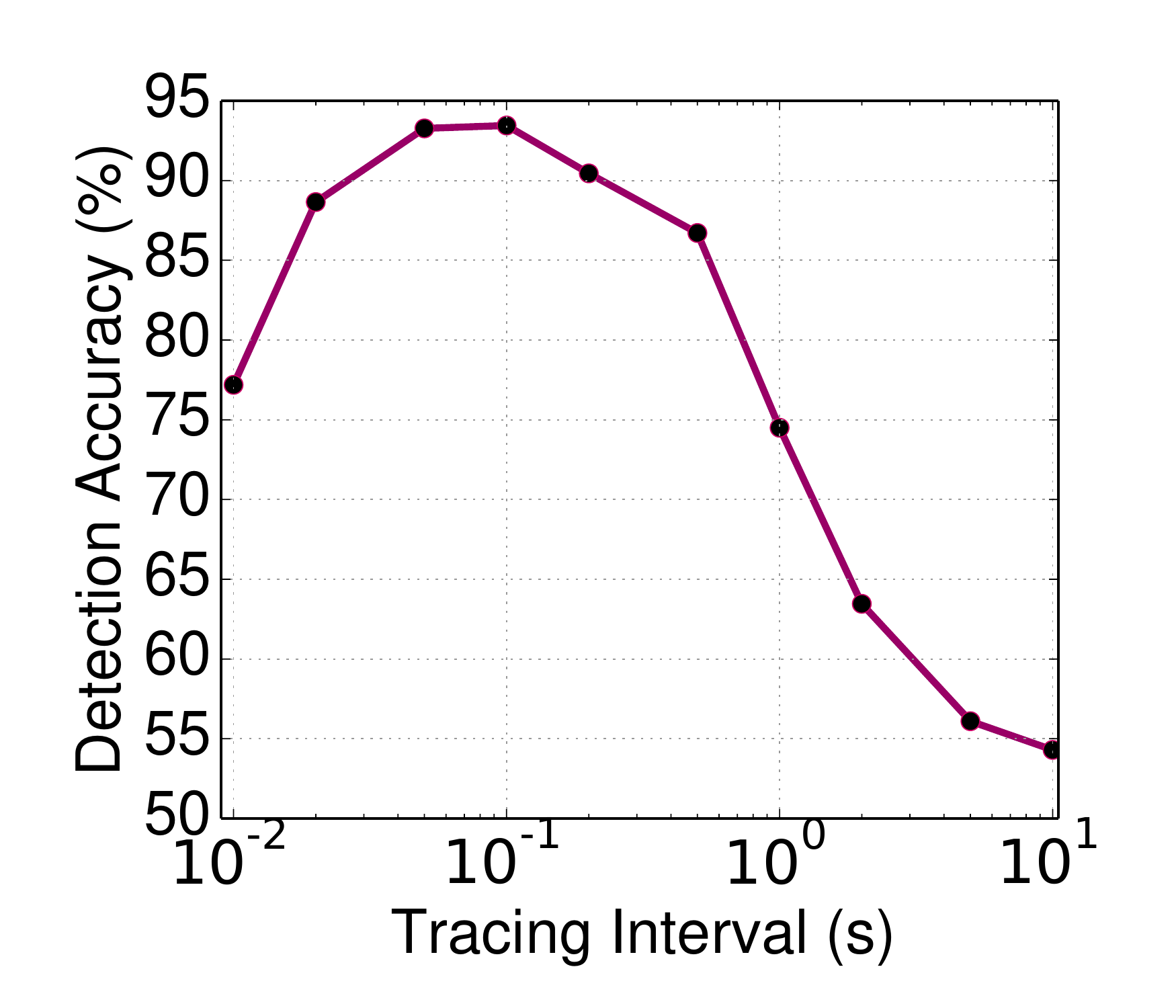}
	\end{tabular}
	\caption{\label{fig:training} Seer's sensitivity to (a) the size of training datasets, and (b) the tracing interval. }
\end{figure}

\noindent{\bf{Sensitivity to training data: }}Fig.~\ref{fig:training}a shows the detection accuracy and training time for Seer as we increase the size of the training dataset. 
The size of the dots is a function of the dataset size. Training data is collected from the 20-server cluster described above, across different load levels, placement strategies, 
time intervals, and request types. The smallest training set size ({\smallcapital 100MB}) is collected over ten minutes of the cluster operating at high utilization, while the largest %training 
dataset ({\smallcapital 1TB}) is collected over almost two months of continuous deployment. As datasets grow Seer's accuracy increases, leveling off at {\smallcapital 100-200GB}. Beyond that 
point accuracy does not further increase, while the time needed for training grows significantly. Unless otherwise specified, we use the 100GB training dataset. % in this section. 

\vspace{0.02in}
\noindent{\bf{Sensitivity to tracing frequency: }}By default the distributed tracing system instantaneously collects the latency of every single user request. 
Collecting queue depth statistics, on the other hand, is a per-microservice iterative process. Fig.~\ref{fig:training}b shows how Seer's accuracy changes as we 
vary the frequency with which we collect queue depth statistics. 
Waiting for a long time before sampling queues, e.g., $>1s$, can result in undetected QoS violations before Seer gets a chance to process the incoming traces. 
In contrast, sampling queues very frequently results in unnecessarily many inferences, and runs the risk of increasing the tracing overhead. 
For the remainder of this work, we use $100ms$ as the interval for measuring queue depths across microservices. 

\begin{figure}
	\centering
	\begin{tabular}{cc}
		& \includegraphics[scale=0.24, viewport = 20 0 400 100]{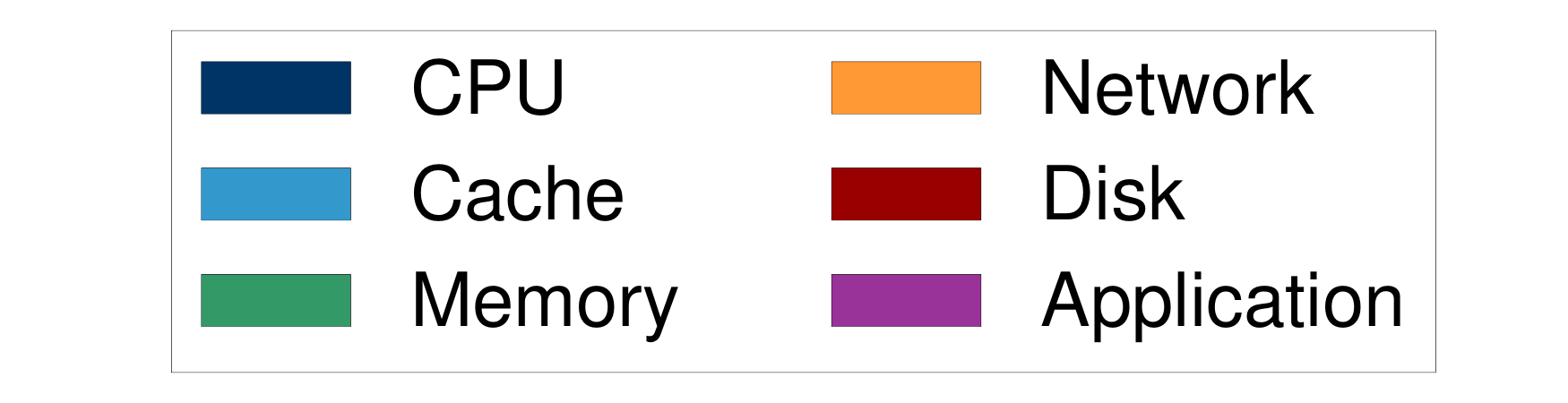} \\
		\includegraphics[scale=0.228, viewport = 115 -5 465 322]{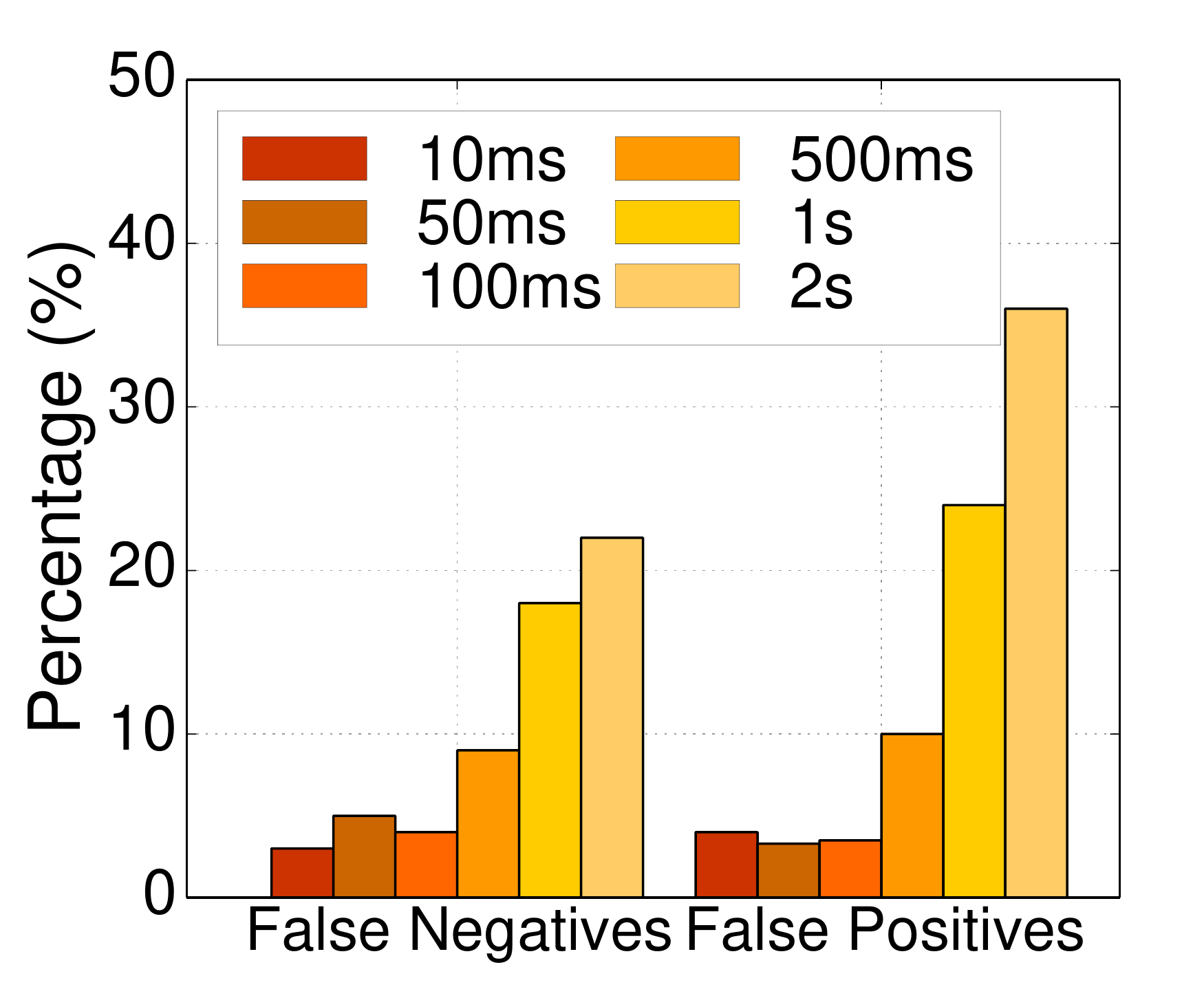} & 
		\includegraphics[scale=0.24, viewport = 0 30 440 420]{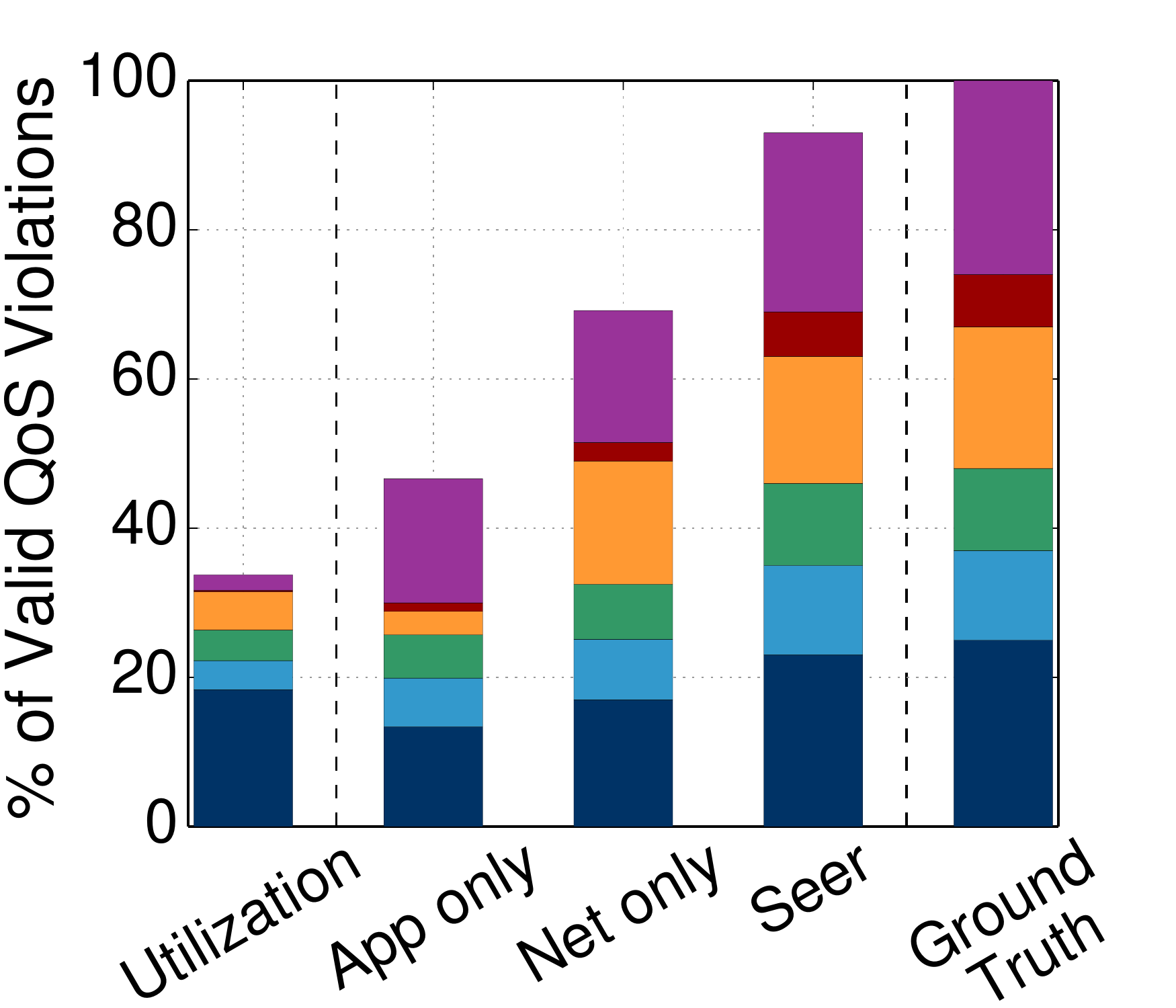} 
	\end{tabular}
		\caption{\label{fig:breakdown} (a) The false negatives and false positives in Seer as we vary the inference window. (b) Breakdown to causes of QoS violations, and comparison with utilization-based detection, and 
		systems with limited instrumentation. }
\end{figure}

\vspace{0.02in}
\noindent{\bf{False negatives \& false positives: }}Fig.~\ref{fig:breakdown}a shows the percentage 
of false negatives and false positives as we vary the prediction window. When Seer tries to anticipate 
QoS violations that will occur in the next 10-100ms both false positives and false negatives are low, 
since Seer uses a very recent snapshot of the cluster state to anticipate performance unpredictability. 
%there is a considerable fraction of false negatives, due to 
If inference was instantaneous, very short prediction windows would always be better. 
However, given that inference takes several milliseconds and more importantly, 
applying corrective actions to avoid QoS violations takes 10-100s of milliseconds to take effect, 
such short windows defy the point of proactive QoS violation detection. At the other end, predicting 
far into the future results in significant false negatives, and especially false positives. This is because 
many QoS violations are caused by very short, bursty events that do not have an impact on queue lengths until a few 
milliseconds before the violation occurs. Therefore requiring Seer to predict one or more seconds into the future means 
that normal queue depths are annotated as QoS violations, resulting in many false positives. % in practice. 
%Similarly, false negatives are caused by Seer not pinpointing the correct source of the QoS violation, which has not yet occurred in the system. 
Unless otherwise specified we use a 100ms prediction window. % for the remainder of the paper. 
%\noindent{\bf{Accuracy: }}Fig.~\ref{fig:accuracy}a shows the detection accuracy in Seer under different input metrics. CPU utilization and per-microservice latencies miss the majority of QoS violations and mislabel the 
%microservices initiating the violations. Using the rate at which per-microservice latency changes achieves higher accuracy but still misses a significant fraction of QoS violations, and incorrectly labels a disproportionate fraction 
%of microservices as culprits. Using the per-microservice queue depth as the input of the neural network captures the majority of QoS violations, and pinpoints the responsible microservice(s). 

%\noindent{\bf{Inference time: }}Fig.~\ref{fig:accuracy}b shows the convergence time for inference in the small 10-server cluster. For 60\% of QoS violations, detection happens within 2ms from obtaining the per-microservice traces, 
%early enough to apply most corrective actions that avoid the QoS violation altogether. Even the QoS violations in the high percentiles of the CDF are detected within 14ms at most, which is usually sufficient for the system to react. %While the cluster remains small, convergence times are within the window of corrective actions in all cases. 

\vspace{0.02in}
\noindent{\bf{Comparison of debugging systems: }}Fig.~\ref{fig:breakdown}b compares Seer 
with a utilization-based performance debugging system that uses resource saturation 
as the trigger to signal a QoS violation, and two systems that only use a fraction of 
Seer's instrumentation. \texttt{App-only} exclusively uses queues measured via 
application instrumentation (not network queues), while \texttt{Network-only} uses 
queues in the {\smallcapital NIC}, and ignores application-level queueing. We also 
show the ground truth for the total number of upcoming QoS violations (96 over a two-week period), 
and break it down by the reason that led to unpredictable performance. 

A large fraction of QoS violations are due to application-level inefficiencies, including correctness bugs, unnecessary synchronization and/or blocking behavior between microservices (including two cases of deadlocks), 
and misconfigured iptables rules, which caused packet drops. An equally large fraction of QoS violations are due to compute contention, followed by contention in the network, cache and memory contention, and finally disk. 
Since the only persistent microservices are the back-end databases, it is reasonable that disk accounts for a small fraction of overall QoS violations. 

Seer accurately follows this breakdown for the most part, only missing a few QoS violations 
due to random load spikes, including one caused by a switch failure. 
The \texttt{App-only} system correctly identifies application-level sources of unpredictable performance, 
but misses the majority of system-related issues, especially in uncore resources. On the other hand, 
\texttt{Network-only} correctly identifies the vast majority of network-related issues, as well as most of 
the core- and uncore-driven QoS violations, but misses several application-level issues. %violations. %sources of unpredictability. 
The difference between \texttt{Network-only} and Seer is small, suggesting that one could omit 
the application-level instrumentation in favor of a simpler design. 
While this system is still effective in capturing %the majority of 
QoS violations, it is less useful in providing feedback to application developers 
on how to improve their design to avoid QoS violations in the future. 
Finally, the utilization-based system behaves the worst, 
missing most violations not caused by {\smallcapital CPU} saturation. % being saturated. 

Out of the 89 QoS violations Seer detects, it notifies the cluster manager early 
enough to avoid 84 of them. The QoS violations that were not avoided 
correspond to application-level bugs, which cannot be easily corrected online. 
Since this is a private cluster, Seer uses utilization metrics and performance counters 
to identify problematic resources. 
%- resolution

%- input and inference, time to convergence

%\noindent{\bf{Scalability: }}We now examine Seer as the cluster size increases. 
%Fig.~\ref{fig:scalability}a shows QoS detection accuracy for different cluster sizes; the 1 and 10 server settings are
%local, while the 40- and 200-instance clusters are on GCE. Seer is robust to the size of the cluster 
%in terms of detection accuracy, although as seen in Fig.~\ref{fig:scalability}b for the 200-instance cluster, 
%the penalty of scalability comes in terms of inference time. For the majority of cases inference takes several hundreds of ms, 
%at which point the QoS violation has occurred. To address this we rewrote Seer using Tensorflow and 
%ported it on the Google TPU public cloud. The change in inference time is dramatic, two orders of magnitude in some cases, 
%ensuring that detection happens early enough to be meaningful. 

\begin{figure}
	\centering
	\includegraphics[scale=0.25, viewport = 340 0 840 100]{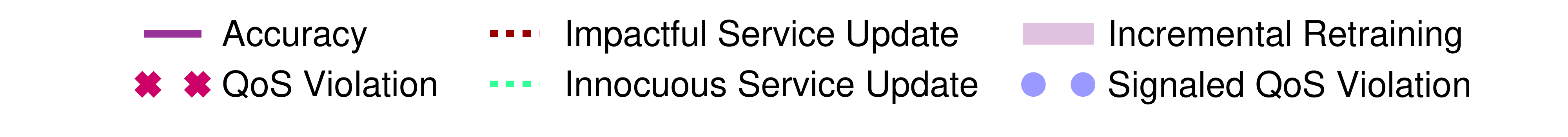} \\
	\includegraphics[scale=0.25, viewport = 205 0 905 300]{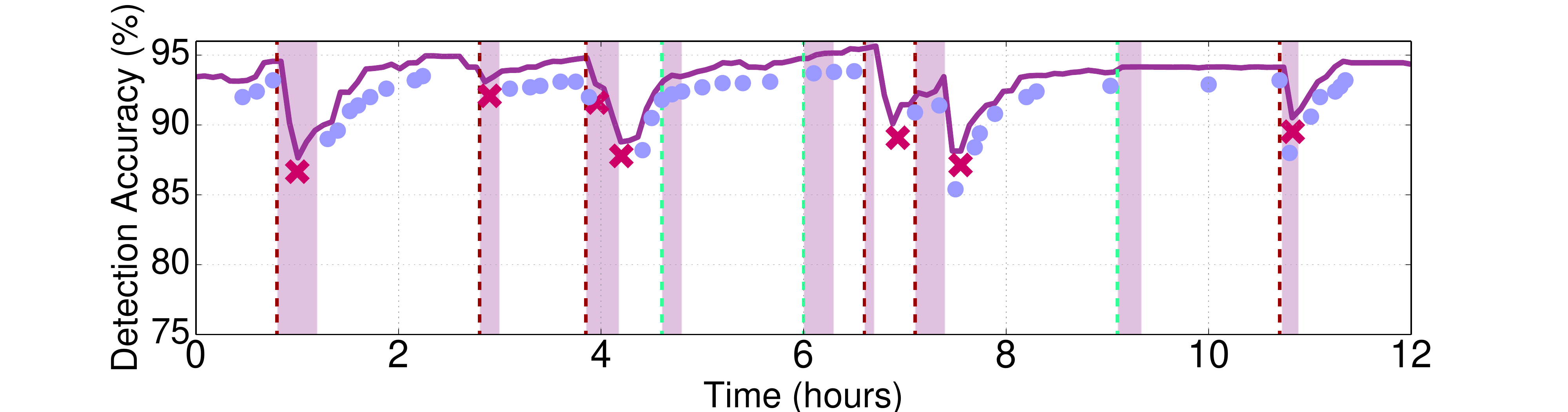} \\
	\includegraphics[scale=0.24, viewport = 340 0 840 80]{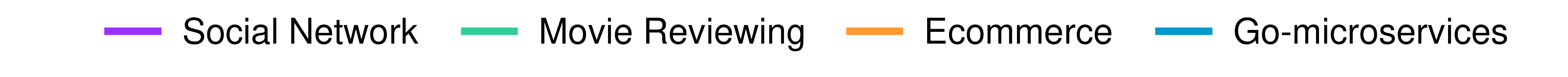} \\
	\includegraphics[scale=0.25, viewport = 205 0 905 300]{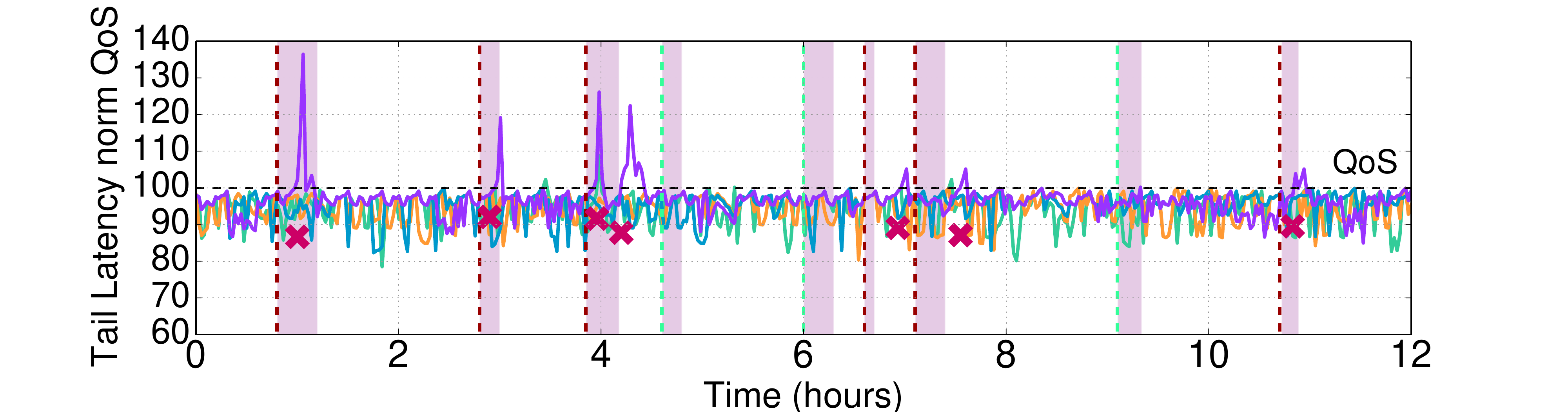}
	\caption{\label{fig:retraining} Seer retraining incrementally after each time the \textit{Social Network} service is updated. }
\end{figure}

\vspace{0.02in}
\noindent{\bf{Retraining: }}Fig.~\ref{fig:retraining} shows the detection accuracy for Seer, and the tail latency for each end-to-end service, over a period of time during which \textit{Social Network} is 
getting frequently and substantially updated. This includes new microservices being added to the service, such as the ability to place an order from an ad using the \texttt{orders} microservice of 
\textit{E-commerce}, or the back-end of the service changing from {\smallcapital\texttt{MongoDB}} to {\smallcapital\texttt{Cassandra}}, and the front-end switching from {\smallcapital\texttt{nginx}} to 
the node.js front-end of \textit{E-commerce}. These are changes that fundamentally affect the application's behavior, throughput, latency, and bottlenecks. The other services remain unchanged during this 
period (\textit{Banking} was not active during this time, and is omitted from the graph). Blue dots denote correctly-signaled upcoming QoS violations, and red $\times$ denote QoS violations that were not 
detected by Seer. All unidentified QoS violations coincide with the application being updated. Shortly after the update Seer incrementally retrains in the background, and starts recovering its 
accuracy until another major update occurs. Some of the updates have no impact on either performance or Seer's accuracy, 
either because they involve microservices off the critical path, or because they are insensitive to resource contention. 

The bottom figure shows that unidentified QoS violations indeed result in performance degradation for \textit{Social Network}, 
and in some cases for the other end-to-end services, if they are sharing physical resources with \textit{Social Network}, 
on an oversubscribed server. Once retraining completes the performance of the service(s) recovers. 
The longer Seer trains on an evolving application, the more likely it is to correctly anticipate its future QoS violations.

%\vspace{-0.06in}
%- initial implementation not fast enough, use TPUs

%initial implementation local, not fast enough, then using TPUs

%tracing 

%training vs. testing

%streaming traces

%governor that signals QoS violation

%agent per-machine to collect hardware stats

%- in public cloud no access to hardware stats, use microbenchmarks instead

%- Results on accuracy

%- TPUs

\section{Large-Scale Cloud Study}
\label{sec:cloud_study}

%Seer does not require dedicating NIC queues to specific microservices, however, dedicated queues reduces contention across network traffic of different microservices. Fig.~\ref{fig:nic} shows 
%the difference between dedicating and load balancing requests across queues for the small-scale cluster. Show number of violations, and how many are detected by Seer. No noticeable difference. 

%- sensitivity to number of neurons, hidden layers, amount of training data

%\begin{figure}
%\centering
%\begin{tabular}{cc}
%\includegraphics[scale=0.26, viewport = 55 0 400 420]{figures/TPU.pdf} &
%\includegraphics[scale=0.26, viewport = 5 0 455 420]{figures/SensitivityTracing.pdf}
%\end{tabular}
%\caption{\label{fig:tpu_brainwave} (a) Seer training and inference with hardware acceleration. (b) Distribution of microservices responsible for QoS violations. } 
%\end{figure}

%- time to infer as a function of training data

%- time to retrain with frequent updates

\subsection{Seer Scalability}

We now deploy our \textit{Social Network} service on a 100-server dedicated cluster on Google Compute Engine (GCE), and use it to service real user traffic. 
The application has 582 registered users, with 165 daily active users, and has been deployed for a two-month period. The cluster on average hosts 
386 single-concerned containers (one microservice per container), subject to some resource scaling actions by the cluster manager, based on Seer's feedback. 

\begin{wrapfigure}[14]{r}{0.28\textwidth}
\centering
\includegraphics[scale=0.29, viewport = 85 0 440 400]{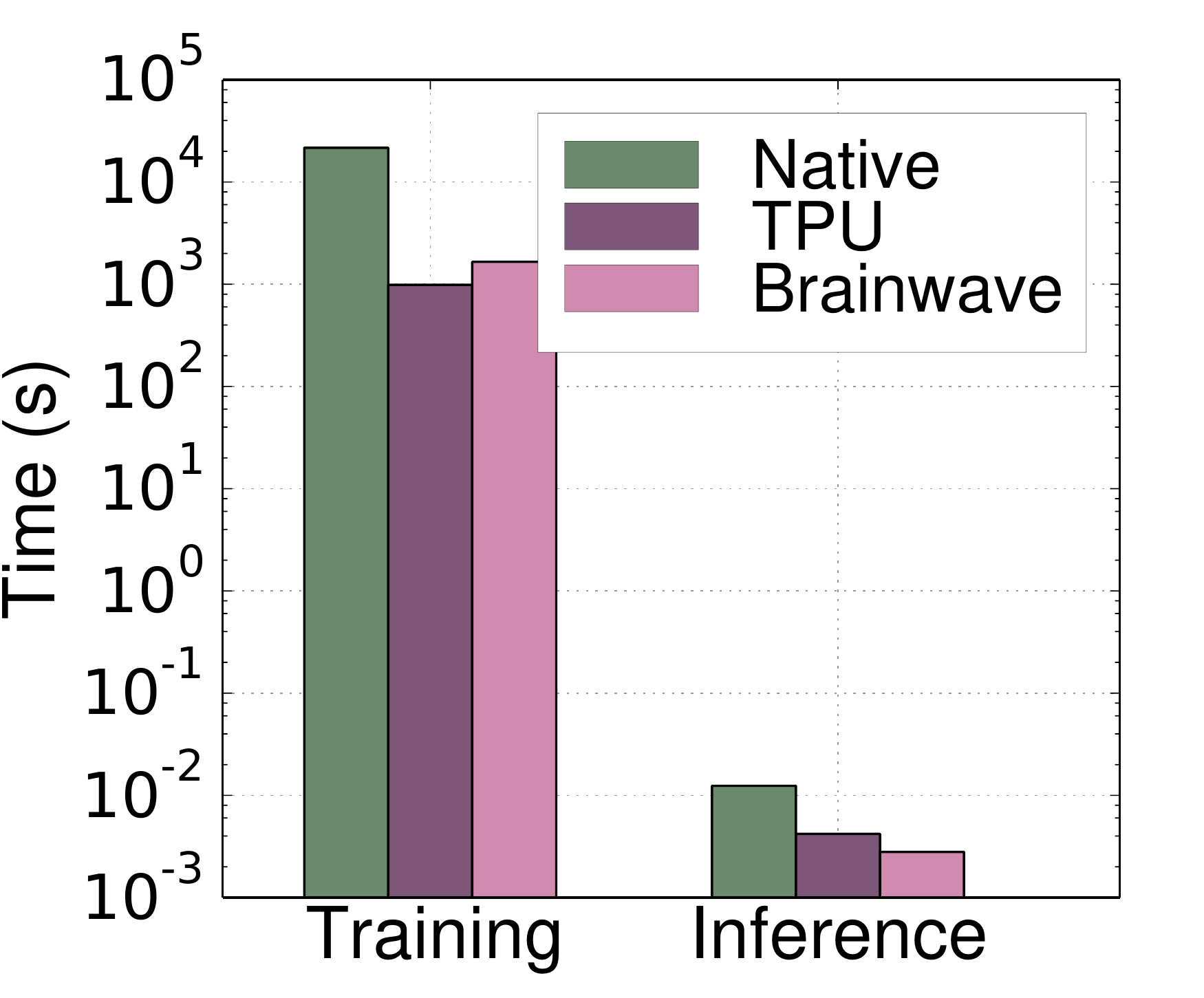}
\vspace{-0.08in}
\caption{\label{fig:tpu_brainwave} Seer training and inference with hardware acceleration. } 
\end{wrapfigure}

Accuracy remains high for Seer, consistent with the small-scale experiments. 
Inference time, however, increases substantially from 11.4ms for the 
20-server cluster to 54ms for the 100-server GCE setting. 
Even though this is still sufficient for many resource allocation decisions, 
as the application scales further, Seer's ability to anticipate a QoS violation 
within the cluster manager's window of opportunity diminishes. 

Over the past year multiple public cloud providers have exposed hardware acceleration offerings 
for {\smallcapital DNN} training and inference, either using a special-purpose design like 
the Tensor Processing Unit ({\smallcapital TPU}) from Google~\cite{tpu}, or using 
reconfigurable {\smallcapital FPGA}s, like Project Brainwave from Microsoft~\cite{brainwave}. 
We offload Seer's {\smallcapital DNN} logic to both systems, and quantify the impact on training and inference time, 
and detection accuracy~\footnote{Before running on TPUs, we reimplemented our DNN in Tensorflow. We similarly adjust the DNN to 
the currently-supported designs in Brainwave. }. Fig.~\ref{fig:tpu_brainwave} shows this comparison 
for a 200GB training dataset. Both the {\smallcapital TPU} and Project Brainwave dramatically 
outperform our local implementation, by up to two orders of magnitude. Between the two 
accelerators, the {\smallcapital TPU} is more effective in training, consistent 
with its design objective~\cite{tpu}, while Project Brainwave achieves faster inference. 
For the remainder of the paper, we run Seer on {\smallcapital TPU}s, and host the \textit{Social Network} service on {\smallcapital GCE}.

\begin{figure}
\centering
\includegraphics[scale=0.26, viewport = 185 10 730 420]{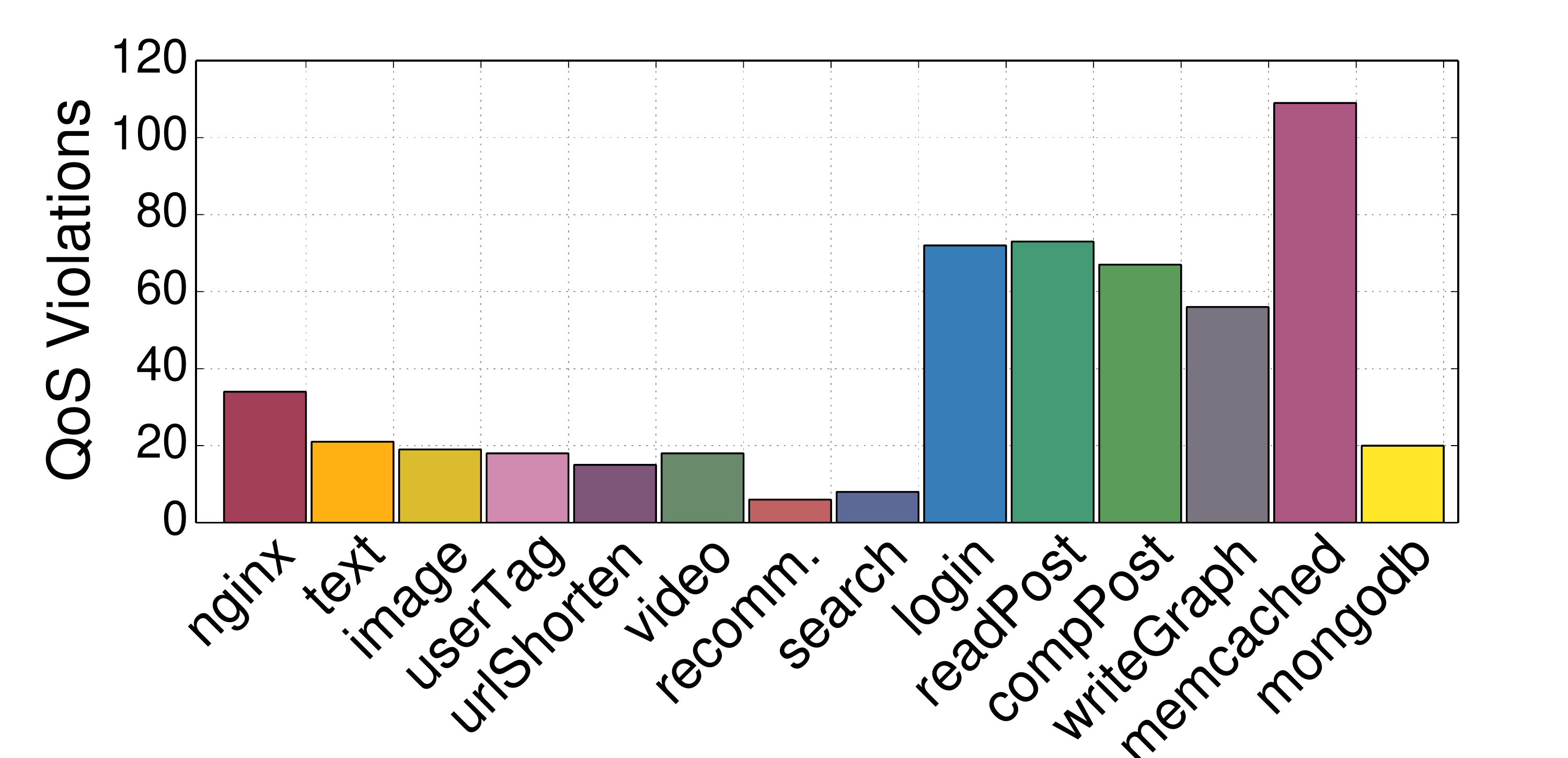}
\caption{\label{fig:culprit} QoS violations each microservice in \textit{Social Network} is responsible for. } 
\end{figure}

\subsection{Source of QoS Violations}

%- distribution across microservices, and why

We now examine which microservice is the most common culprit for a QoS violation. Fig.~\ref{fig:culprit} shows the number of QoS violations caused by each service over the two-month period. The
most frequent culprits by far are the in-memory caching tiers in {\smallcapital\texttt{memcached}}, and Thrift services with high request fanout, such as {\smallcapital\texttt{composePost}}, 
{\smallcapital\texttt{readPost}}, and {\smallcapital\texttt{login}}. {\smallcapital\texttt{memcached}} is a justified source of QoS violations, since it is on the critical path for almost all query types, 
and it is additionally very sensitive to resource contention in compute and to a lesser degree cache and memory. Microservices with high fanout are also expected to initiate QoS violations, 
as they have to synchronize multiple inbound requests before proceeding. If processing for any incoming requests is delayed, end-to-end performance is likely to suffer. %it will affect %The Thrift services with high fanout are also 
Among these QoS violations, most of {\smallcapital{\texttt{memcached}}}'s violations were caused by resource contention, while violations in Thrift services were caused by long synchronization times. 
%{\smallcapital{\texttt{}}}

%Over the two month period, Seer has detected

%\subsection{QoS Violation Prevention}

%- per-machine diagnostics, how many QoS violations are avoided (not just detected)

%- private vs. public cloud (large scale EC2 cluster from before)

\subsection{Seer's Long-Term Impact on Application Design} %Diagnostics}

\begin{figure}
\centering
\includegraphics[scale=0.28, viewport = 185 20 730 360]{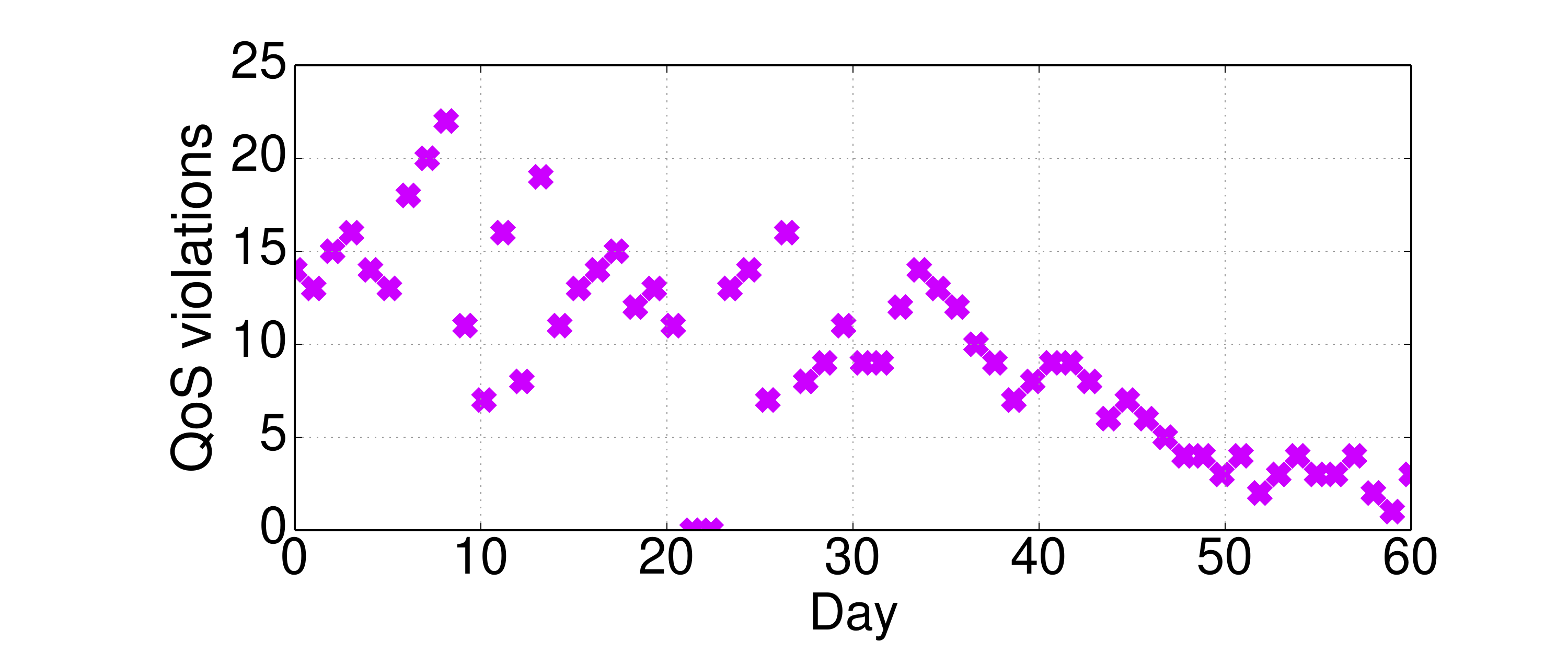}
\caption{\label{fig:errors_time} QoS violations each microservice in \textit{Social Network} is responsible for. } 
\end{figure}

Seer has now been deployed in the \textit{Social Network} cluster for over two months, 
and in this time it has detected 536 upcoming QoS violations (90.6\% accuracy) and avoided 495 (84\%) of them. 
Furthermore, by detecting recurring patterns that lead to QoS violations, 
Seer has helped the application developers better understand bugs and design decisions 
that lead to hotspots, such as microservices with a lot of back-and-forth communication 
between them, or microservices forming cyclic dependencies, or using blocking primitives. 
This has led to a decreasing number of QoS violations over the two month period (seen in 
Fig.~\ref{fig:errors_time}), as the application progressively improves. In days 22 and 23 
there was a cluster outage, which is why the reported violations are zero. Systems like 
Seer can be used not only to improve performance predictability in complex cloud systems, 
but to help users better understand the design challenges of microservices, 
as more services transition to this application model. 

\vspace{-0.06in}
\section{Conclusions}

Cloud services increasingly move away from complex monolithic designs, 
and adopt the model of specialized, loosely-coupled microservices. 
%Cloud systems and applications continuously increase in size and complexity. The recent switch from monoliths to microservices 
%puts even more pressure on performance predictability, and at the same time makes manual performance debugging impractical. 
We presented Seer, a data-driven cloud performance debugging system 
that leverages practical learning techniques, and the massive amount of tracing 
data cloud systems collect to proactively detect and avoid QoS violations. 
We have validated Seer's accuracy in controlled environments, 
and evaluated its scalability on large-scale clusters on public clouds. 
We have also deployed the system in a cluster hosting a social network 
with hundreds of users. In all scenarios, Seer accurately detects 
upcoming QoS violations, improving responsiveness and performance 
predictability. As more services transition to the microservices model, 
systems like Seer provide practical solutions that can %effectively 
navigate the increasing complexity of the cloud. 

\clearpage
\balance
%
% The next two lines define the bibliography style to be used, and the bibliography file.
\bibliographystyle{ACM-Reference-Format}
\bibliography{references}

\end{document}